\documentclass{emulateapj}
\usepackage{graphicx}

\shorttitle{PULSAR KICKS AND SPIN TILTS IN CLOSE DOUBLE NEUTRON STARS} 
\shortauthors{Willems, Kalogera, \& Henninger}

\begin{document}

\title{PULSAR KICKS AND SPIN TILTS IN THE CLOSE DOUBLE NEUTRON STARS
  PSR\,J0737-3039, PSR\,B1534+12 AND PSR\,B1913+16}

\author{B. Willems, V. Kalogera, and M. Henninger}
\affil{Northwestern University, Department of Physics and Astronomy,
  2145 Sheridan Road, Evanston, IL 60208, USA}
\email{b-willems, vicky@northwestern.edu, \\
m-henninger@alumni.northwestern.edu} 

\begin{abstract}
In view of the recent measurement of the scintillation velocity for
PSR\,J0737-3039, we examine the complete set of constraints imposed on
the pulsar B natal kicks (magnitude and orientation) and predict the
most favorable pulsar kick velocity and spin tilt for both isotropic
and polar kicks. Our analysis takes into account both currently
unknown parameters: the orientation of the orbital plane on the sky
($\Omega$) and the radial component of the systemic velocity
($V_{r}$). Assuming that the system's peculiar velocity is entirely
due to the second supernova explosion, we find that the system may
have crossed the Galactic plane multiple times since the birth of the
second neutron star and that the post-supernova peculiar velocity
could have been as high as $\simeq 1200$\,km\,s$^{-1}$. We also
confirm the absolute lower and upper limits on the physical parameters
derived in our earlier study. For specific combinations of the two
unknown parameters $\Omega$ and $V_r$, however, we find much tighter
constraints on the pre-supernova binary configuration and natal kicks
imparted to pulsar B, as well as on the age of system. Once $\Omega$
is measured in the coming year, it will be straightforward to use the
results presented here to further constrain the natal kicks and the
spin-tilt predictions. We complete our comprehensive study and derive
similar constraints and spin-tilt predictions for PSR\,B1534+12, where
the only free parameter is $V_{r}$. Lastly, for PSR\,B1913+16, we
update the progenitor and kick constraints using the measured pulsar
spin tilt and allowing for Roche-lobe overflow from the progenitor of
the pulsar companion.
\end{abstract}

\keywords{Stars: Binaries: Close, Stars: Pulsars: General, Stars:
  Neutron, Stars: Pulsars: Individual: (PSR\,J0737-3039,
  PSR\,B1534+12, PSR\,B1913+16)} 

\section{Introduction}

The recent discovery of a unique, strongly relativistic binary pulsar
(Burgay et al. 2003) that is also the first eclipsing, double pulsar
system found in our Galaxy (Lyne et al. 2004) promises to provide us
with more impressive tests of general relativity than ever before
(Kramer 2004), to greatly improve our understanding of magnetospheric
and pulsar emission physics (Lyutikov 2004; McLaughlin et al. 2004;
Demorest et al. 2004; Granot \& M\'esz\'aros 2004; Arons et al. 2004),
and possibly indicates a realistic chance for the detection of
gravitational waves from double-neutron-star (DNS) inspiral with the
first-generation of ground-based interferometers (Kalogera et
al. 2004; Kim et al. 2004).

Apart from the binary pulsar and eclipsing behavior, PSR\,0737-3039
sets the current records for DNS properties: for example, it has the
shortest spin and orbital period, smallest eccentricity, lowest
magnetic field, and fastest apsidal motion. The measured current
orbital parameters and masses along with an estimate of the system's
age have allowed the careful account of its recent evolutionary
history, since the time just before the supernova (SN) that formed
pulsar B, the current companion to the millisecond pulsar A (Willems
\& Kalogera 2004; Dewi \& van den Heuvel 2004). In Willems \& Kalogera
(2004) (hereafter
Paper I), we found that the pre-SN binary\footnote{Throughout the
paper, we use the terms pre- and post-SN to refer to the instants just
before and right after the SN explosion forming the second NS.} most
likely consisted of the first-formed NS and a Roche-lobe-filling
helium star with mass between $2.1\,M_\odot$ and $4.7\,M_\odot$ (cf.\
Dewi \& van den Heuvel 2004). The pre-SN orbital separation can be
constrained between $1.36\, R_\odot$ and $1.72\, R_\odot$, and the
magnitude of the kick velocity between 60\,km\,s$^{-1}$ and
1560\,km\,s$^{-1}$.

Most recently, scintillation observations of pulsar A have allowed the
measurement of the two systemic velocity components in the reference
frame of the orbital plane projected on the sky (Ransom et al.\
2004). This measurement gives us the first glimpse of the kinematic
evolutionary history (in addition to that related to binary
evolution). Therefore we extend the analysis presented in Paper I and
take a comprehensive approach to constraining the formation of
PSR\,J0737-3039. The concept is similar to that adopted by Wex,
Kalogera, \& Kramer (2000) with two main differences: (i) at present
there is no constraint on pulsar A's spin tilt from timing or
polarization observations; (ii) the current kinematic measurements
involve two free parameters, the orientation of the orbital plane on
the sky ($\Omega$) and the radial component of the systemic velocity
($V_{r}$). 
\begin{deluxetable*}{llccc}
\tabletypesize{\scriptsize}
\tablewidth{0pt}
\tablecolumns{5}
\tablecaption{Parameters of PSR\,J0737-3039, PSR\,B1534+12, and
  PSR\,B1913+16. \label{param}} 
\tablehead{ \colhead{Parameter} & 
     \colhead{Notation} & 
     \colhead{PSR\,J0737-3039\tablenotemark{a}} &
     \colhead{PSR\,B1534+12\tablenotemark{b}} &
     \colhead{PSR\,B1913+16\tablenotemark{c}}
     } 
\startdata
Right Ascension (J2000)             & $\alpha$     & $07^{\rm h} 37^{\rm m} 51.25^{\rm s}$            
                                                   & $15^{\rm h} 37^{\rm m} 09.96^{\rm s}$ 
                                                   & $19^{\rm h} 15^{\rm m} 28.00^{\rm s}$ \\
Declination (J2000)                 & $\delta$     & $-30^\circ 39^\prime 40.74^{\prime \prime}$                       
                                                   & $11^\circ 55^\prime 55.55^{\prime \prime}$ 
                                                   & $16^\circ 06^\prime 27.40^{\prime \prime}$ \\
Proper motion in R.A.               & $\mu_\alpha$ (mas\,yr$^{-1}$) & - &   $1.34 \pm 0.01$ & $-3.27 \pm 0.35$ \\
Proper motion in Decl.              & $\mu_\delta$ (mas\,yr$^{-1}$) & - & $-25.05 \pm 0.02$ & $-1.04 \pm 0.42$ \\
Galactic longitude (J2000)          & $l$          & $245^\circ\!\!\!.\,2$ 
                                                   & $19^\circ\!\!\!.\,8$
                                                   & $50^\circ\!\!\!.\,0$ \\
Galactic latitude (J2000)           & $b$          & $ -4^\circ\!\!\!.\,5$ 
                                                   & $48^\circ\!\!\!.\,3$ 
                                                   & $2^\circ\!\!\!.\,1$ \\
Proper motion in Gal.\ long.        & $\mu_l$ (mas\,yr$^{-1}$) & - & -21.57\tablenotemark{d} & -0.60 \\
Proper motion in Gal.\ lat.         & $\mu_b$ (mas\,yr$^{-1}$) & - & -12.80\tablenotemark{d} & -3.38 \\
Distance                            & $d$ (kpc)    & $\sim 0.6$ & $1.02 \pm 0.05$ & $8.3 \pm 1.4$ \\
Characteristic age of the ms pulsar & $\tau_c$ (Myr) & 210 & 250 & 110 \\
Spin-down age of the ms pulsar      & $\tau_b$ (Myr) & 100 & 210 & 80 \\
Mass of the ms pulsar               & $M_A$ ($M_\odot$) & 1.34 & 1.33 & 1.44 \\
Mass of the companion               & $M_B$ ($M_\odot$) & 1.25 & 1.35 & 1.39 \\
Current semi-major axis             & $A_{cur}$ ($R_\odot$) & 1.26 & 3.28 & 2.80 \\
Current orbital eccentricity        & $e_{cur}$    & 0.0878 & 0.274 & 0.617 
\enddata
\tablenotetext{a}{Burgay et al. 2003; Lyne et al. 2004.}
\tablenotetext{b}{Wolszczan 1991; Stairs et al. 2002; Konacki,
  Wolszczan, \& Stairs 2003; Arzoumanian, Cordes, \& Wasserman 1999.} 
\tablenotetext{c}{Hulse \& Taylor 1975; Taylor et al. 1976; Taylor,
  Fowler, \& McCulloch 1979; Taylor \& Weisberg 1982, 1989; Damour \&
  Taylor 1991; Arzoumanian et al. 1999, Wex et al. 2000.} 
\tablenotetext{d}{Derived quantities.}
\end{deluxetable*}
In this paper, we account for the complete set of equations
describing the formation of PSR\,J0737-3039 and derive constraints on
the progenitor of pulsar B and on the kicks (both isotropic and polar)
imparted to it at birth. We also derive predictions for the possible
orientation of pulsar~A's spin axis with respect to the current
orbital angular momentum axis. We carefully analyze and present the
dependence of these results on $\Omega$ and $V_{r}$ and on the
system's age. The measurement of $\Omega$ along with proper motion
components is anticipated in the coming year. We show that such
measurements can be used with the results presented here to derive
tighter constraints.  We also compare our predicted spin tilts with
those derived based on the geometric model of Jenet \& Ransom (2004)
for pulsar B's eclipses and find excellent agreement. We note that our
predicted spin tilts can also be used along with polarization
observations to anticipate the most likely timing of pulsar A's
disappearance due to geodetic precession as a function of the radial
velocity (which in the long term will be the only remaining unknown
property) of the system.

In addition to PSR\,J0737-3039 (Section~\ref{sec0737}), we analyze our
current understanding of the evolutionary history of the other two
close currently known DNS systems in the Galaxy. For PSR\,B1534+12, we
use the masses, orbital characteristics, and measured proper motion to
derive progenitor and kick constraints, as well as spin tilt
predictions (Section~\ref{sec1534}). For PSR\,B1913+16, we extend the
analysis of Wex et al.\ (2000) to take into account the possibility
that the helium star progenitor of the pulsar companion could have
been filling its Roche lobe. We also derive the probability
distributions of the kick magnitude imparted to the last-born NS
(Section~\ref{sec1913}). We end with a summary of our main
conclusions, a discussion of the implications of our analysis for the
relation of kicks to spin orientations, and a comparison of our
results with other recent studies.

\section{THE RECENT EVOLUTIONARY HISTORY OF PSR\,J0737-3039}  
\label{sec0737}

In this Section, we use the scintillation velocities measured by
Ransom et al. (2004) to first determine the spatial velocity of
PSR\,J0737-3039 with respect to a frame of reference locked to the
Galaxy and next follow the motion of the system in the Galaxy
backwards in time until the time of its formation. We identify
crossings of the Galactic plane as possible birth sites and, by
subtracting the local Galactic rotational velocity from the total
systemic velocity at birth, determine the system's post-SN peculiar
velocity. This velocity is subsequently used to tighten the progenitor
constraints derived in Paper~I.

The physical parameters of PSR\,J0737-3039 relevant to this
investigation are summarized in Table~\ref{param}. We particularly
note that we use the estimate $\tau_b = 100$\,Myr of the time since
pulsar~A left the maximum spin-up line as an {\em upper limit} for the
age of the system (for details, see Paper~I).

\subsection{Present Systemic Velocity}

The interstellar scintillation measurements carried out by Ransom et
al.\ (2004) yield two of the three systemic velocity components
required to describe the space motion of PSR\,J0737-3039. Both
components are measured in the plane perpendicular to the
line-of-sight and with respect to the Earth. The first one has a
magnitude of $\simeq 96.6 \pm 3.7\, {\rm km\,s^{-1}}$ and is directed
along the line connecting the ascending and descending node of the
pulsar's orbit, while the second one has a magnitude of $\simeq 103.1
\pm 7.7\, {\rm km\,s^{-1}}$ and is directed perpendicular to the nodal
line. In what follows, we will refer to these two components as
$V^\parallel$ and $V^\perp$, respectively. Since the statistical
errors of $3.7\, {\rm km\,s^{-1}}$ and $7.7\, {\rm km\,s^{-1}}$ do not
account for unknown systematic errors, we here adopt the values
$V^\parallel = 100\, {\rm km\,s^{-1}}$ and $V^\perp = 110 {\rm
km\,s^{-1}}$, and assume an uncertainty on both components of $\pm 20
{\rm km\,s^{-1}}$ (Ransom 2004, private communication).

Before we pass on from the scintillation velocity components to velocity
components with respect to a frame of reference locked to the Galaxy,
it is interesting to examine the velocity components $V_\alpha$ and
$V_\delta$, and the corresponding proper motion components
$\mu_\alpha$ and $\mu_\delta$, in right ascension ($\alpha$) and
declination ($\delta$). The velocity components $V_\alpha$ and
$V_\delta$ are obtained from $V^\parallel$ and $V^\perp$ by means of
the transformation formulae
\begin{equation}
\renewcommand{\arraystretch}{1.5}
\left.
\begin{array}{lcl}
V_\alpha & = & V^\parallel\, \cos \Omega 
   - V^\perp\, \sin \Omega,  \nonumber \\
V_\delta & = & V^\parallel\, \sin \Omega  
   + V^\perp\, \cos \Omega,  \nonumber 
\end{array}
\right\} \label{pm1}
\end{equation}
where $\Omega$ is the {\it unknown} longitude of the ascending node
measured right-handed around the line-of-sight, from the East through
the North. The proper motion components $\mu_\alpha = \cos \delta\,
d\alpha/dt$ and $\mu_\delta = d\delta/dt$ are then given by the
relations
\begin{equation}
\renewcommand{\arraystretch}{2.3}
\left.
\begin{array}{lcl}
\displaystyle
{V_\alpha \over {\rm km\,s^{-1}}} & = & 4.74\, 
   \displaystyle \left( {d \over {\rm kpc}} \right)
   \left( {\mu_\alpha \over {\rm mas\,yr^{-1}}} \right),
   \nonumber \\
\displaystyle
{V_\delta \over {\rm km\,s^{-1}}} & = & 4.74\, 
   \displaystyle \left( {d \over {\rm kpc}} \right)
   \left( {\mu_\delta \over {\rm mas\,yr^{-1}}} \right),
   \nonumber 
\end{array}
\right\} \label{pm2}  
\end{equation}
where $d$ is the distance of the binary from the Sun. 

\begin{figure}
\resizebox{\hsize}{!}{\includegraphics{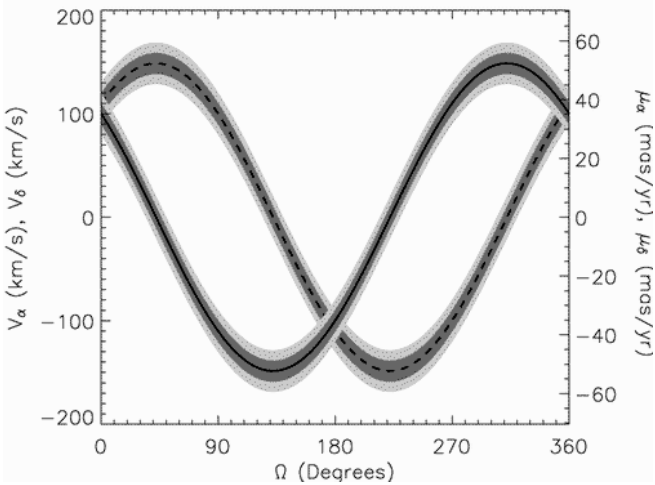}}
\caption{Variation of the velocity components of PSR\,J0737-3039 in
  right ascension $\alpha$ (solid lines) and declination $\delta$
  (dashed lines) as a function of the unknown longitude of the
  ascending node $\Omega$. The dark and light gray bands indicate the
  errors in $V_\alpha$ and $V_\delta$ corresponding to errors of 10
  and 20\,km\,s$^{-1}$ in the scintillation velocity components
  $V^\parallel$ and $V^\perp$. The right-hand axis shows the magnitude
  of the proper motion components $\mu_\alpha$ and $\mu_\delta$, for a
  distance of 0.6\,kpc. }
\label{vad}
\end{figure}

The variations of $V_\alpha$ (or $\mu_\alpha$) and $V_\delta$ (or
$\mu_\delta$) are shown in Fig.\,\ref{vad} as a function of the
unknown $\Omega$. The dark and light gray bands indicate the errors in
$V_\alpha$ (or $\mu_\alpha$) and $V_\delta$ (or $\mu_\delta$)
corresponding to errors of 10 and 20\,km\,s$^{-1}$ in $V^\parallel$
and $V^\perp$. The magnitude of $V_\alpha$ and $V_\delta$ ranges from
approximately $-175\, {\rm km\,s^{-1}}$ to $175\, {\rm km\,s^{-1}}$,
while, for a distance of 0.6\,kpc, the magnitude of $\mu_\alpha$ and
$\mu_\delta$ can be anywhere between $-60\, {\rm mas\,yr^{-1}}$ and
$60\, {\rm mas\,yr^{-1}}$ (see also Ransom et al. 2004). A 100\,mas
proper motion should then be detectable in less than 2 years time. It
is furthermore clear that, even with the extended error bars of $\pm
20\, {\rm km\,s^{-1}}$, future measurements of the proper motion will
considerably constrain the possible values of the longitude of the
ascending node $\Omega$.

Since the scintillation velocity measurements only give the magnitude
of $V^\parallel$ and $V^\perp$, the motion of the binary projected on
the plane perpendicular to the line-of-sight should, in principal, be
considered in four different directions corresponding to four
different combinations of $V^\parallel = \pm 100\, {\rm km\,s^{-1}}$
and $V^\perp = \pm 110 {\rm km\,s^{-1}}$. This additional degree of
freedom may however be incorporated in the unknown longitude of the
ascending node. To illustrate this, we consider two sets of
scintillation velocity components $( V^\parallel_1, V^\perp_1 )$ and
$( V^\parallel_2, V^\perp_2 )$ whose orientations in the plane
perpendicular to the line-of-sight are determined by the angles
$\Omega_1$ and $\Omega_2$. From Eqs.~(\ref{pm1}), it then follows that
the two sets of scintillation velocity components give rise to equal
velocity components in right ascension and declination when $\Omega_2
= \Omega_1 + \Delta \Omega$, where $\Delta \Omega$ is determined by
\begin{equation}
\renewcommand{\arraystretch}{3.0}
\left.
\begin{array}{lcl}
\displaystyle \cos (\Delta \Omega) = 
   {V_1^\parallel \over V_2^\parallel} + 
   {V_2^\perp \over V_2^\parallel}\, {{V_1^\perp\, V_2^\parallel - 
   V_1^\parallel\, V_2^\perp} \over 
   {( V_2^\parallel )^2 + 
   ( V_2^\perp )^2}},  \nonumber \\
\displaystyle \sin (\Delta \Omega) = {{V_1^\perp\, V_2^\parallel - 
   V_1^\parallel\, V_2^\perp} \over 
   {( V_2^\parallel )^2 + 
   ( V_2^\perp )^2}}.  \nonumber 
\end{array}
\right\} \label{pm6}
\end{equation}
Hence, the cases $\left( V^\parallel, V^\perp \right) = \left( -100\, 
{\rm km\,s^{-1}}, 110\, {\rm km\,s^{-1}} \right)$, $\left(
V^\parallel, V^\perp \right) = \left( 100\, {\rm km\,s^{-1}}, -110\,
{\rm km\,s^{-1}} \right)$, and $\left( V^\parallel, V^\perp \right)
= \left( -100\, {\rm km\,s^{-1}}, -110\, {\rm km\,s^{-1}} \right)$,
can be obtained from the case $\left( V^\parallel, V^\perp \right) =
\left( 100\, {\rm km\,s^{-1}}, 110\, {\rm km\,s^{-1}} \right)$ by
translating $\Omega$ over angles of $275.5^\circ$, $95.5^\circ$,
and $180^\circ$, respectively. For the remainder of the paper, we
therefore restrict ourselves to presenting the results associated
with the velocity components $V^\parallel = 100\, {\rm km\,s^{-1}}$
and $V^\perp = 110\, {\rm km\,s^{-1}}$.

\subsection{Kinematic History and Age}
\label{motion}

In order to trace the Galactic motion of PSR\,J0737-3039 back in time
until the time of its formation, we first need to determine the three
velocity components of the current systemic velocity in a reference
frame locked to the Galaxy. For this purpose, we choose a right-handed
frame $OXYZ$ with the origin at the Galactic center and with the
$XY$-plane coinciding with the Galactic plane. We adopt the direction
from the Sun to the Galactic center as the positive direction of the
$X$-axis and the direction of the Sun's motion due to the Galactic
rotation as the positive direction of the $Y$-axis.

The components $(V_X,V_Y,V_Z)$ of the current systemic velocity with
respect to $OXYZ$ are obtained by the vector transformation
\begin{equation}
\left( 
   \begin{array}{c}
   V_X \\ V_Y \\ V_Z
   \end{array} 
\right)
= \vec{V}_\odot + J \cdot
\left( 
   \begin{array}{c}
   V_r \\ V_\alpha \\ V_\delta
   \end{array} 
\right),  \label{uvw}
\end{equation}
where $\vec{V}_\odot$ is the Galactic velocity of the Sun, $J=\partial
(X, Y, Z)/\partial (d, \alpha, \delta)$ is the Jacobian of the
coordinate transformation linking $(X, Y, Z)$ to $(d, \alpha,
\delta)$, and $V_r$ is the {\it unknown} radial velocity. The Galactic
velocity of the Sun in turn is the vector sum of the Sun's circular
velocity around the Galactic center and the Sun's peculiar velocity
with respect to the local standard of rest:
\begin{equation}
\vec{V}_\odot
= 
\left( 
   \begin{array}{c}
   0\, {\rm km\,s^{-1}} \\ 221\, {\rm km\,s^{-1}} \\ 
   0\, {\rm km\,s^{-1}}
   \end{array} 
\right) + \left( 
   \begin{array}{c}
   10\, {\rm km\,s^{-1}} \\ 5\, {\rm km\,s^{-1}} \\ 
   7\, {\rm km\,s^{-1}}
   \end{array} 
\right),  \label{vsun}
\end{equation}
(Binney \& Dehnen 1997, Bienaym\'e 1999). It is evident from
Eqs.~(\ref{pm1}) and~(\ref{uvw}) that the kinematic evolution of
PSR\,J0737-3039 depends on two unknown parameters: $\Omega$ and $V_r$.

\begin{figure}
\resizebox{\hsize}{!}{\includegraphics{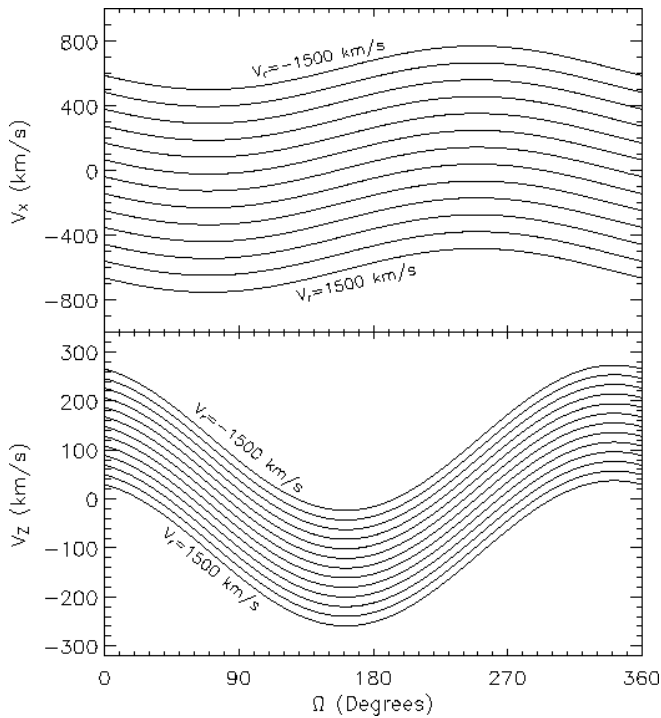}}
\caption{Variation of the velocity components $V_X$ and $V_Z$ of
  PSR\,J0737-3039 as a function of the unknown longitude
  of the ascending node $\Omega$, for $V_r$-values
  ranging from -1500 to 1500\,km\,s$^{-1}$ in increments of
  250\,km\,s$^{-1}$. } 
\label{vxvz}
\end{figure}

The variations of $V_X$ and $V_Z$ as functions of $\Omega$ are
displayed in Fig.~\ref{vxvz}, for $V_r = -1500$, -1250, $-1000,
\ldots$, 1500\,km\,s$^{-1}$. It is interesting to note that for $V_r
\la -250$\,km\,s$^{-1}$, $V_X$ is always positive so that the motion
of the system backwards in time is away from the Galactic center. In
addition, for $\Omega \la 25^\circ$ and $\Omega \ga 300^\circ$, $V_Z$
is positive for all considered values of $V_r$ so that the motion of
the system backwards in time is also away from the Galactic plane. As
we will see below, this behaviour together with the current position
of the system below the Galactic plane has important consequences for
the past motion of PSR\,J0737-3039 in the Galactic potential and the
identification of possible birth places.

Given the present position and velocity components of PSR\,J0737-3039
in the adopted reference frame, we integrate the equations governing
the system's motion in the Galaxy backwards in time and look for
possible birth sites as a function of $\Omega$ and $V_r$. Since DNS
systems are thought to form from primordial binaries with initially
two massive component stars, which are known to have a very small
vertical scale height of 50--70\,pc, we assume the primordial binary
to be born in the Galactic disk. Wex et al. (2000) furthermore noted
that both the orbital velocity before the birth of the first NS and
the center-of-mass velocity imparted to the system due to the first SN
explosion are typically of the order of only a few 10\,km\,s$^{-1}$
(see also Brandt \& Podsiadlowski 1995; Pfahl et al. 2002). However,
Pfahl et al. (2002) used a population synthesis technique to construct
a probability distribution of post-SN peculiar velocities of high-mass
X-ray binaries and found a low-probability tail extending all the way
up to $\simeq 200$\,km\,s$^{-1}$. The possibility that the first SN
explosion in PSR\,J0737-3039 produced a kick of a few
100\,km\,s$^{-1}$ to the binary's center-of-mass can therefore not be
entirely ruled out. In order to properly treat these cases, the effect
of both the first and second SN explosion on the space motion of the
binary must be taken into account, which requires the introduction of
three additional free parameters (magnitude of the first SN kick,
direction of the first SN kick relative to that of the second kick,
and time expired between the two kicks). In view of this added
complexity and the low probability of the high systemic velocities
after the first SN explosion, we here leave aside these more
complicated and rarer cases and assume the present kinematic
properties of PSR\,J0737-3039 to be entirely due to the second SN
explosion. In particular, we assume that the system was still close to
the Galactic plane at the time of the second SN explosion and that its
pre-SN {\em peculiar} velocity was small in comparison to the Galactic
rotational velocity at the formation site. The post-SN systemic
velocity of the binary is then given by the sum of the local Galactic
rotational velocity and the peculiar velocity due to the second SN
explosion.

Following Wex et al. (2000), we use the Galactic potential of Kuijken
\& Gilmore (1989) to calculate the motion of PSR\,J0737-3039 in the
Galaxy backwards in time for all possible $\Omega$-values between
$0^\circ$ and $360^\circ$, and $V_r$-values between -1500 and
1500\,km\,s$^{-1}$. Each time the binary crosses the Galactic plane
(Z=0), we check whether the crossing occurred less than 100\,Myr in
the past and whether the crossing point is within 15\,kpc from the
Galactic center. If so, the crossing is considered as a possible birth
site and the post-SN peculiar velocity $V_{\rm pec}$ is determined by
subtracting the local Galactic rotational velocity from the system's
total systemic velocity\footnote{Note that Table~3 in Wex et
al. (2000) contained an error in the coefficients $\beta_1$ for
$A=2,3$; where the correct values are $\beta_1=0$ for both $A=2$ and
3. We corrected this error and confirmed that it does not
quantitatively affect the results presented by Wex at al. (2000) for
PSR\,B1913+16 (see also Section~\ref{sec1913}). In particular, the
resulting relative errors on the times of the Galactic plane crossings
and the associated peculiar velocities are less than $\sim 6\%$.}.

\begin{figure}
\epsscale{.80}
\plotone{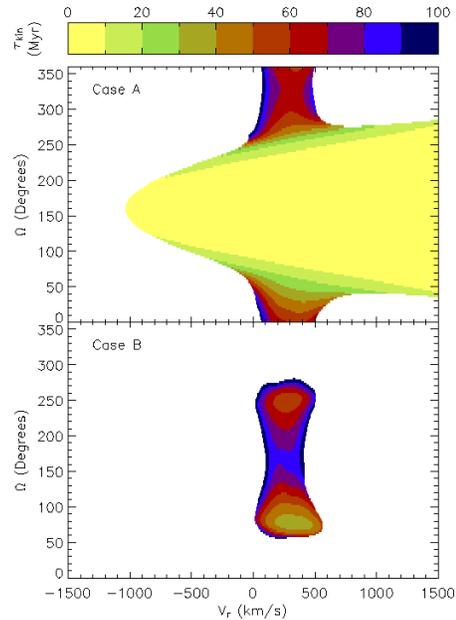}
\caption{Kinematic age of PSR\,J0737-3039 as a function of the unknown
  radial velocity $V_r$ and the unknown longitude of the ascending
  node $\Omega$, for both case~A (top panel) and~B (bottom panel).}  
\label{tauk1}
\end{figure} 

\begin{figure*}
\resizebox{\hsize}{!}{\includegraphics{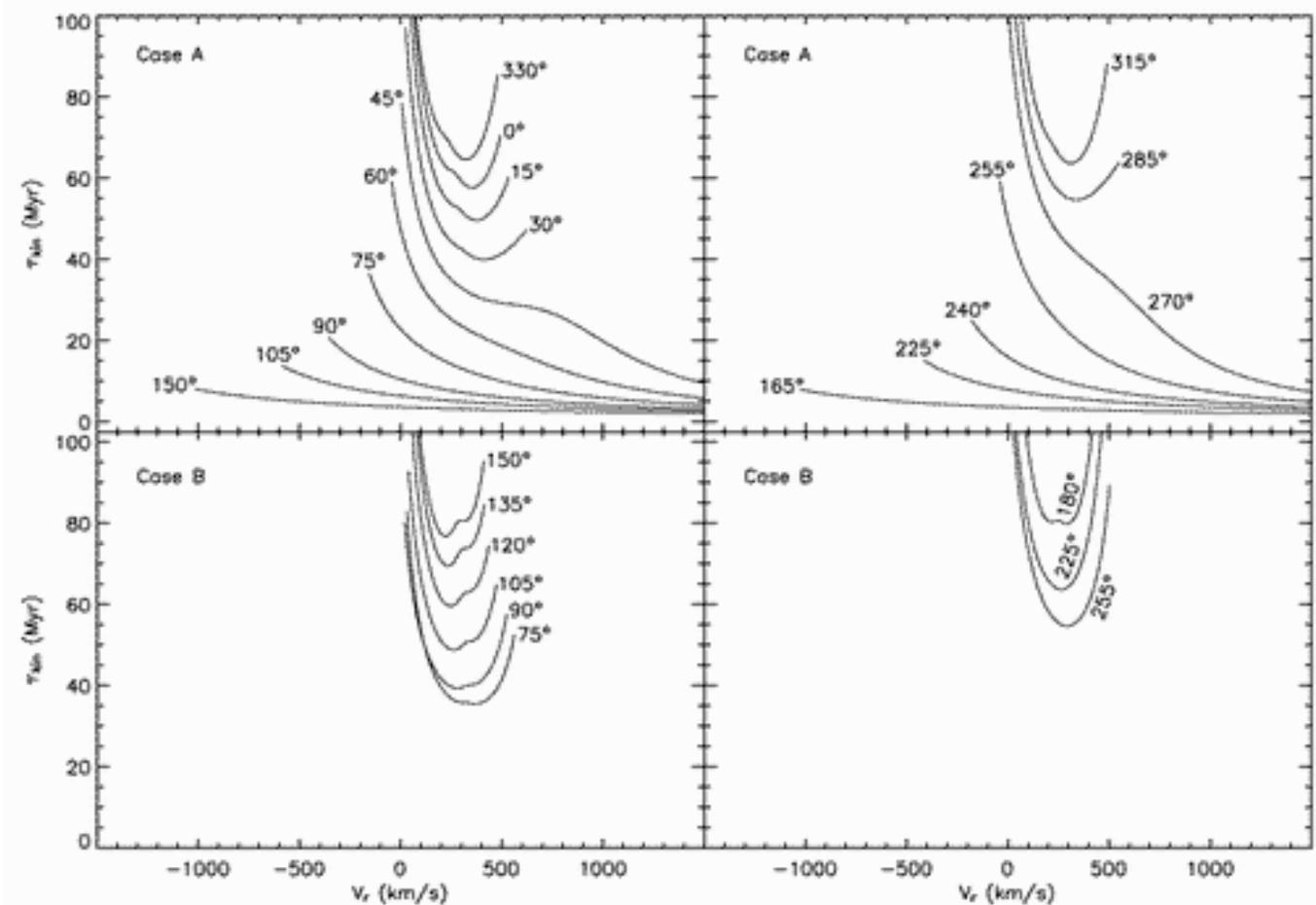}}
\caption{Cross sections of Fig.~\ref{tauk1} showing the variation of
  the kinematic age $\tau_{kin}$ of PSR\,J0737-3039 as a function of
  $V_r$, for different values of $\Omega$. The top and bottom panels
  correspond to case~A and~B, respectively.}
\label{tauk3}
\end{figure*}

In cases where the binary crosses the Galactic plane within 1\,kpc
from the Galactic center, we assume that the pre-SN binary belonged to
the disk-like component of the Galactic bulge (see, e.g., Kuijken \&
Rich 2002). Since the time required by a massive binary to evolve into
a DNS is at most of the order of $\sim 20$\,Myr, much smaller than the
typical age of 'pure' Galactic bulge stars, we exclude birth sites in
the bulge at $Z \ne 0$. Hence, all disk crossings within 100\,Myr in
the past and within 15\,kpc from the Galactic center are considered as
possible birth sites and are assumed to follow the disk-like rotation
curve determined by the adopted Galactic potential.

The time in the past at which the system crosses the Galactic disk
provides an estimate for the time since the formation of the
DNS. Since some of the Galactic trajectories cross the Galactic plane
more than once, these kinematic ages are not uniquely determined by
$\Omega$ and $V_r$. In particular, for $0\, {\rm km\,s^{-1}} \la V_r
\la 550 {\rm km\,s^{-1}}$ and $60^\circ \la \Omega \la 280^\circ$, we
find that the system may have crossed the Galactic plane twice within
the last 100\,Myr; while for $50\, {\rm km\,s^{-1}} \la V_r \la 400
{\rm km\,s^{-1}}$ and $240^\circ \la \Omega \la 265^\circ$ it may have
even crossed the Galactic plane as much as three times. Since the
latter situation only occurs for just a few, greatly fine-tuned
trajectories at ages very close to the adopted upper limit of
100\,Myr, we leave aside these rather rare cases and focus on the
first and second Galactic plane crossings. We will refer to these
crossings as case~A and case~B, respectively.

The dependence of the kinematic age $\tau_{kin}$ on both $\Omega$ and
$V_r$ is displayed in Figs.~\ref{tauk1} and~\ref{tauk3}:
Fig.~\ref{tauk1} shows the kinematic ages for all considered $\Omega$-
and $V_r$-values using a linear gray scale (color in the electronic
edition), while Fig.~\ref{tauk3} gives a more detailed view of the
dependence of $\tau_{kin}$ on $V_r$ for some specific values of
$\Omega$.

For case~A, the majority of the Galactic disk crossings occur less
than 20\,Myr in the past. These young ages are associated with
longitudes of the ascending node between $\simeq 30^\circ$ and $\simeq
260^\circ$, and radial velocities larger than approximately
-1000\,km\,s$^{-1}$. Galactic disk crossings occurring more than
20\,Myr in the past are typically associated with $\Omega$-values
smaller than $\simeq 70^\circ$ or larger than $\simeq 230^\circ$, and
with $V_r$-values between -200\,km\,s$^{-1}$ and
700\,km\,s$^{-1}$. The range of possible $V_r$-values becomes narrower
with increasing kinematic ages and is most restricted when the age of
the system is close to the adopted upper limit of 100\,Myr. There are
furthermore almost no disk crossings associated with negative
$V_r$-values when $\tau_{kin} \ga 20$\,Myr. The origin of this
systematic behavior is related to the dependence of the $X$- and
$Z$-components of the current peculiar velocity on $V_{r}$ and
$\Omega$ (see Fig.~\ref{vxvz}): the Galactic position of the system is
such that negative $V_{r}$-values correspond to orbits where the first
disk crossing occurs either too far in the past (in excess of even the
characteristic age of $210$\,Myr) or too far away from the Galactic
center (at radii in excess of 15\,kpc). The disk crossings associated
with negative $V_r$-values and $\tau_{kin} \la 20$\,Myr occur due to
the suitable orientation of the orbit in space: the $\Omega$-values
($60^\circ \la \Omega \la 230^\circ$) are such that the $Z$-component
of the current peculiar velocity is negative, so that motion backwards
in time is towards the Galactic plane. In addition, Lorimer et
al. (2004) recently estimated the age of the system to be most likely
between 30\,Myr and 70\,Myr. If the age is indeed so constrained, the
range of possible $\Omega$-values would be limited to $\Omega \la
80^\circ$ and $\Omega \ga 240^\circ$, and the range of possible
$V_r$-values to $-200\,{\rm km\,s^{-1}} \la V_r \la 800\,{\rm
km\,s^{-1}}$. In view of the current uncertainties we will perform our
analysis for the upper limit of 100\,Myr, but the above age
constraints may be kept in mind.

Case~B disk crossings, on the other hand, only occur for $V_r$-values
between 0\,km\,s$^{-1}$ and $550 {\rm km\,s^{-1}}$, and
$\Omega$-values between $60^\circ$ and $280^\circ$. The associated
kinematic ages are always longer than 20\,Myr, so that constraining
the age to 30--70\,Myr does not impose severe additional limits on the
ranges of $\Omega$- and $V_r$-values. 

\begin{figure}
\epsscale{.80}
\plotone{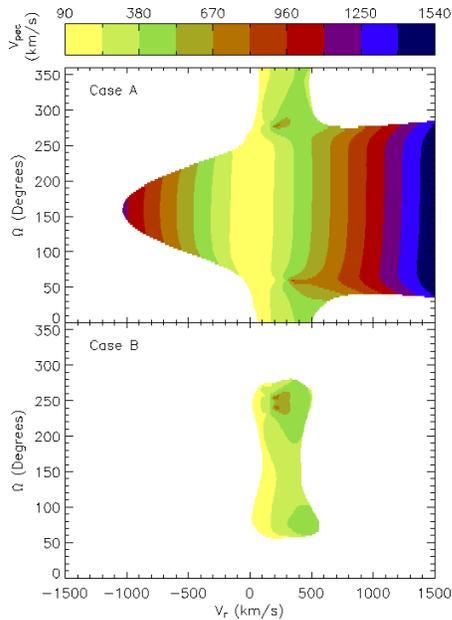}
\caption{Peculiar velocity of PSR\,J0737-3039 as a function of the
  unknown radial velocity $V_r$ and the unknown longitude of the
  ascending node $\Omega$. The top and bottom panels correspond to
  case~A and~B, respectively. }
\label{vcm1}
\end{figure}

In Fig.~\ref{vcm1}, we show the peculiar velocities $V_{\rm pec}$
associated with the Galactic plane crossings at ages shown in
Fig.~\ref{tauk1}. The peculiar velocities range from 90\,km\,s$^{-1}$
to 1550\,km\,s$^{-1}$ and from 120\,km\,s$^{-1}$ to 800\,km\,s$^{-1}$,
for case~A and~B, respectively.  However, when $\tau_{kin} \ga
40$\,Myr, the post-SN peculiar velocities associated with case~A are
also confined to a smaller interval between 90\,km\,s$^{-1}$ and
800\,km\,s$^{-1}$. In addition, for a given value of $\Omega$, the
peculiar velocity $V_{\rm pec}$ generally increases with increasing
absolute values of $V_r$. In Section~\ref{ss0737}, we will use the
post-SN peculiar velocities to further constraint the properties of
the pre-SN progenitor of PSR\,J0737-3039 derived in Paper~I.

We conclude this section by noting that, if PSR\,J0737-3039 is indeed
born in the Galactic disk, Figs.~\ref{tauk1} and~\ref{vcm1} show that
the future determination of $\Omega$ may yield important constraints
on its age and post-SN orbital velocity, even without knowing the
system's present radial velocity.

\subsection{Progenitor Constraints}
\label{ss0737}

In this Section, we extend the analysis presented in Paper~I by adding
the kinematic history of PSR\,J0737-3039 to the constraints resulting
from the orbital dynamics of asymmetric, instantaneous SN explosions.
The two main differences with paper~I are that (i) the problem is now
expressed as a function of the two unknown parameters $\Omega$ and
$V_r$, and (ii) the spin-down age $\tau_b=100\,{\rm Myr}$ is now used
as an upper limit for the age instead of as an actual age
estimate. The first difference implies that the constraints will also
become functions of $\Omega$ and $V_r$, while the second difference
affects the amount of orbital evolution that may have taken place
since the birth of the system.

As in Paper~I, we assume that the post-SN orbit evolved solely under
the influence of gravitational radiation and determine the semi-major
axis $A$ and orbital eccentricity $e$ right after the second SN
explosion by numerically integrating the orbital evolution equations
derived by Junker \& Sch\"{a}fer (1992) backwards in time. For each
combination of $\Omega$ and $V_r$, the integration is terminated at
the kinematic age $\tau_{kin}$ corresponding to a case~A or case~B
Galactic disk crossing. The resulting post-SN orbital separations and
orbital eccentricities range from $A=1.26\, R_\odot$ and $e=0.088$
($\tau_{kin}=0$\,Myr) to $A=1.54\, R_\odot$ and $e=0.12$
($\tau_{kin}=100$\,Myr).

Under the assumption that the pre-SN binary orbit is circular (see
Paper~I), the pre- and post-SN binary parameters are related by the
conservation laws of orbital energy and orbital angular momentum. The
relations take the form
\begin{eqnarray}
\lefteqn{V_k^2 + V_0^2 + 2\, V_k\, V_0\, \cos \theta}
   \nonumber \\
 & & = G \left( M_A + M_B \right) \left( {2 \over A_0} 
 - {1 \over A} \right),  \label{eq1}
\end{eqnarray}
\begin{eqnarray}
\lefteqn{A_0^2 \left[ V_k^2\, \sin^2 \theta\, \cos^2 \phi \right. 
  +  \left. \left( V_k\, \cos \theta + V_0
 \right)^2 \right]}  \nonumber \\
 & & = G \left( M_A + M_B \right) A 
 \left( 1 - e^2 \right) \hspace{1.2cm} \label{eq2}
\end{eqnarray}
(e.g., Hills 1983; Kalogera 1996), where $G$ denotes the gravitational
constant, $A_0$ the pre-SN orbital separation, $M_0$ and $V_0 = \left[
G \left( M_A + M_0 \right)/A_0 \right]^{1/2}$ the helium star's mass
and pre-SN orbital velocity, and $V_k$ the magnitude of the kick
velocity imparted to pulsar~B at birth. The angles $\theta$ and $\phi$
describe the direction of the kick velocity imparted to pulsar~B:
$\theta$ is the polar angle between $\vec{V}_k$ and $\vec{V}_0$, while
$\phi$ is the corresponding azimuthal angle measured in the plane
perpendicular to $\vec{V}_0$ so that $\phi=\pi/2$ coincides with the
direction from pulsar~A to pulsar~B.  For a given kick-velocity
magnitude and a given set of pre- and post-SN binary parameters, the
polar angle $\theta \in [0,\pi]$ is uniquely determined by
Eq.~(\ref{eq1}). The azimuthal angle $\phi \in [0,2\pi]$, on the other
hand, is only determined through the square of its cosine so that, for
a given kick-velocity magnitude, four solutions to Eqs.~(\ref{eq1})
and~(\ref{eq2}) exist for each set of pre- and post-SN binary
parameters.

A necessary set of conditions for the existence of solutions of
Eqs.~(\ref{eq1}) and~(\ref{eq2}) is that $0 \le \sin^2 \theta \le1$
and $0 \le \cos^2 \phi \le 1$. Together with the requirements that the
binary must remain bound after the SN explosion, that the post-SN
orbit must pass through the position of both stars at the time of the
SN explosion, and that the SN kick must reproduce the post-SN orbital
parameters $A$ and $e$, these conditions impose strong constraints on
the possible pre-SN values of $A_0$ and $M_0$ as well as on the
magnitude of the kick velocity $V_k$ that have already been described
in detail in Paper~I (see also Fryer \& Kalogera 1997; Kalogera \&
Lorimer 2000).  In addition to these constraints, the kick velocity
and the pre- and post-SN binary parameters must now also be compatible
with the post-SN peculiar velocity derived by following the motion of
the system in the Galaxy backwards in time. From Eqs.~(3) and~(34) in
Kalogera (1996), it follows that the post-SN peculiar velocity is
related to the kick velocity and to the pre- and post-SN binary
parameters as
\begin{eqnarray}
\lefteqn{V_{\rm pec}^2 = {{M_B\,M_0} \over {(M_A+M_0) (M_A+M_B)}} 
  \left[ {{G (M_0-M_B) M_A} \over {M_0\, A}} \right. }
  \nonumber \\
 & & \left. + {{G (M_0-M_B) (M_0-2\,M_B) M_A} \over 
  {M_0\, M_B\, A_0}} + V_k^2  \right]. \hspace{1.7cm} \label{vpec}
\end{eqnarray}
We require the peculiar velocity obtained from this equation to be
within the error bars of the peculiar velocity obtained by following
the motion of the system backwards in time. In order to estimate the
error bars on the latter, we neglect the contributions resulting from
uncertainties in the adopted Galactic potential and assume them to be
entirely due to the errors in the scintillation velocity measurements
($\pm 20$\,km\,s$^{-1}$).  

\begin{figure*}
\resizebox{\hsize}{!}{\includegraphics{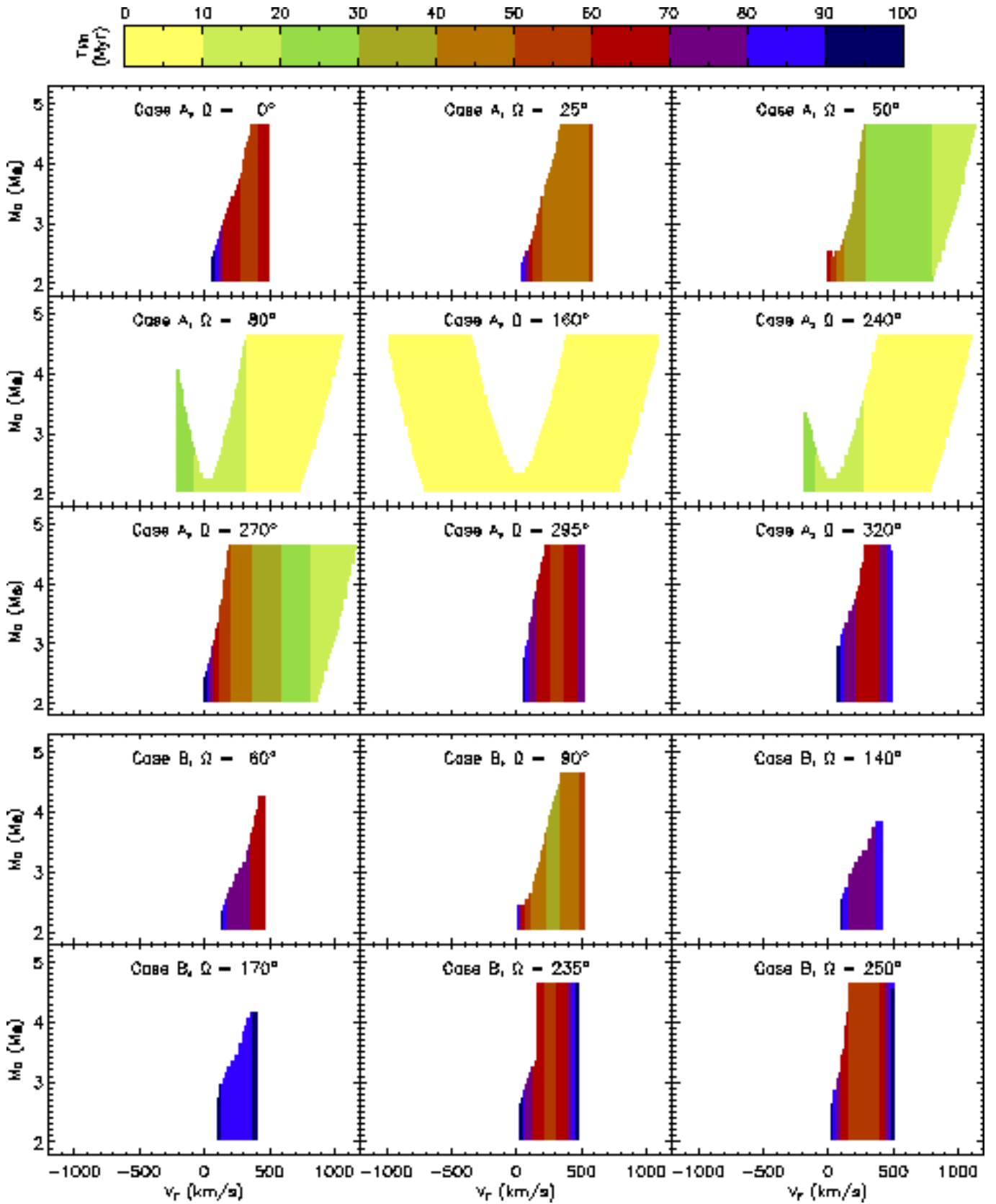}}
\caption{Limits on the mass $M_0$ of PSR\,J0737-3039B's pre-SN helium
  star progenitor as a function of $V_r$, for various values of
  $\Omega$ which were chosen as to provide the clearest and most
  informative graphical representation. The upper 9 panels are for
  case~A; the lower 6 panels for case~B. The gray scale (color in the
  electronic edition) indicates the kinematic ages associated with
  different combinations of $\Omega$ and $V_r$
  (cf. Fig.~\ref{tauk1}).}
\label{pro1}
\end{figure*}

\begin{figure*}
\resizebox{\hsize}{!}{\includegraphics{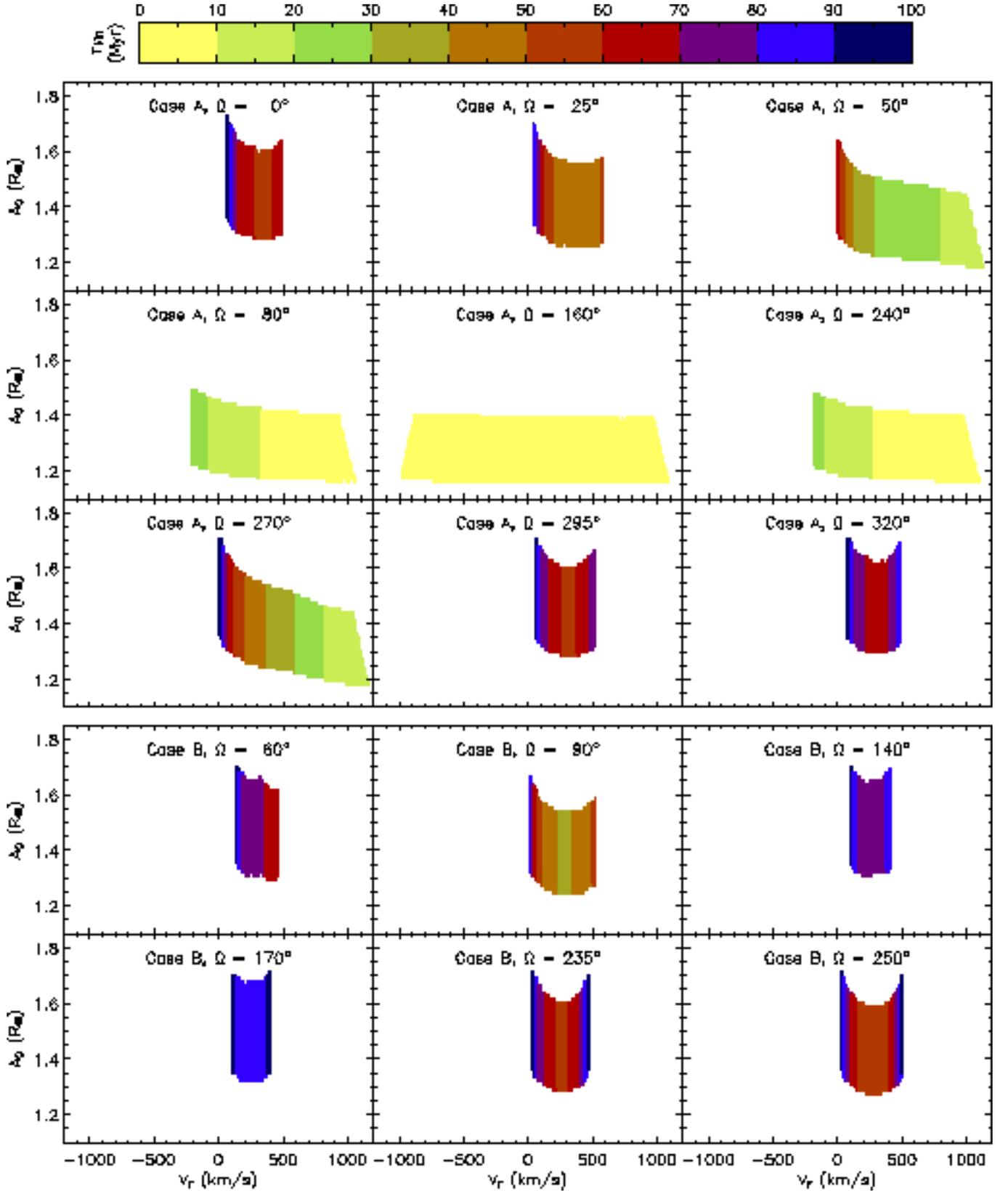}}
\caption{Limits on the pre-SN orbital separation $A_0$ of
  PSR\,J0737-3039 as a function of $V_r$, for various values of
  $\Omega$. The upper 9 panels are for case~A; the lower 6 panels for
  case~B. The gray scale (color in the electronic edition) indicates
  the kinematic ages associated with different combinations of
  $\Omega$ and $V_r$ (cf. Fig.~\ref{tauk1}).}
\label{pro2}
\end{figure*}

\begin{figure*}
\resizebox{\hsize}{!}{\includegraphics{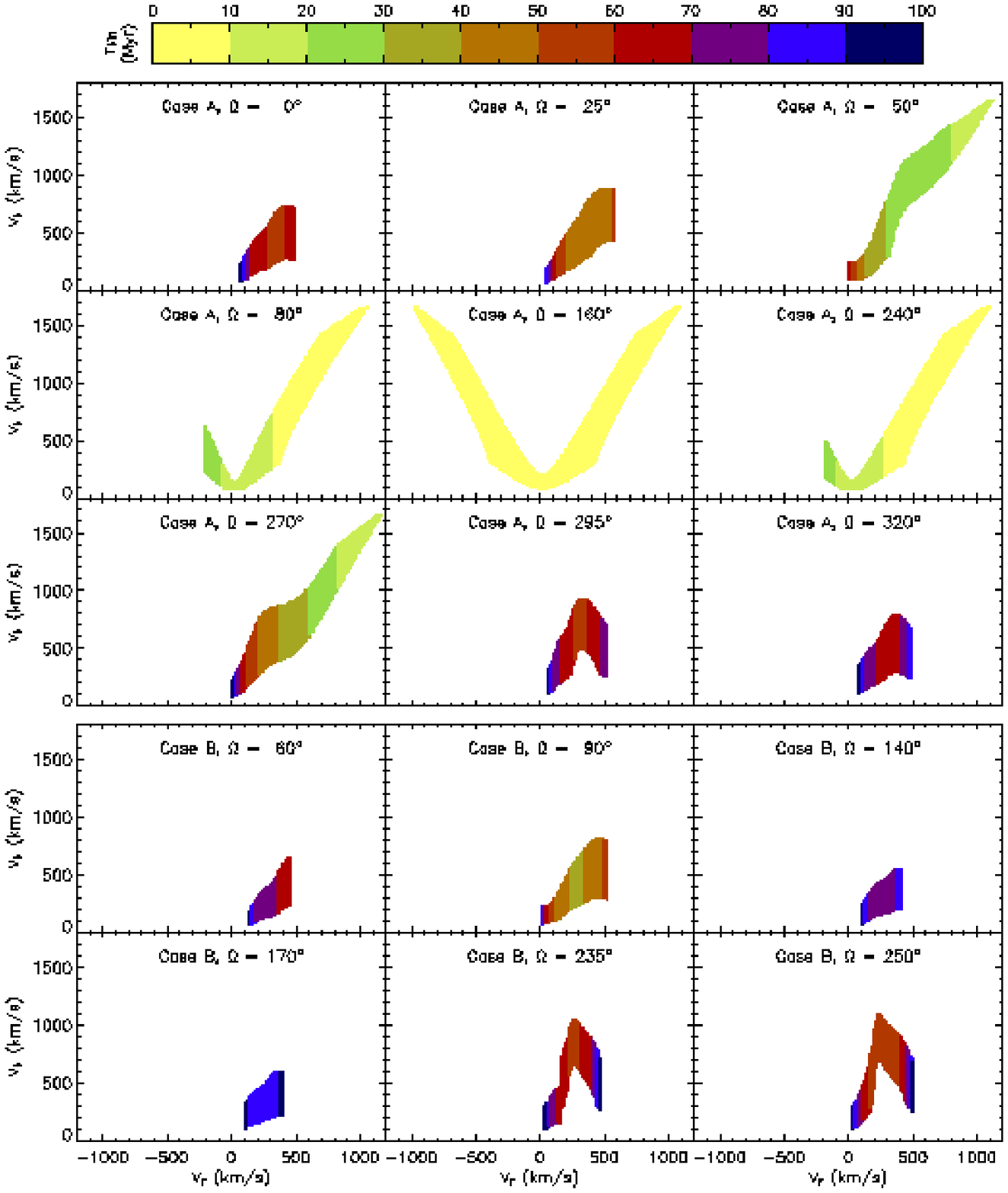}}
\caption{Limits on the magnitude $V_k$ of the kick velocity imparted
  to PSR\,J0737-3039B at its formation as a function of $V_r$, for
  various values of $\Omega$. The upper 9 panels are for case~A; the
  lower 6 panels for case~B. The gray scale (color in the electronic
  edition) indicates the kinematic ages associated with different
  combinations of $\Omega$ and $V_r$ (cf. Fig.~\ref{tauk1}).}
\label{pro3}
\end{figure*}

In Figs.~\ref{pro1}--\ref{pro3}, we show the ranges of $M_0$, $A_0$,
and $V_k$ values satisfying {\it all} the imposed constraints as
functions of the unknown radial velocity $V_r$, for case A and case B,
and for different specific values of the longitude of the ascending
node $\Omega$. The parameter space for case~A disk crossings is
displayed in the upper 9 panels of the figures, whereas the parameter
space for case~B disk crossings is displayed in the lower 6
panels. The gray scale (color in the electronic edition) indicates the
kinematic ages associated with different combinations of $\Omega$ and
$V_r$ (cf. Fig.~\ref{tauk1}).

For each value of $\Omega$, there is a finite range of $V_r$-values
leading to physically acceptable solutions for the pre-SN progenitor
of PSR\,J0737-3039. For case~A disk crossings, the range is widest for
$\Omega \simeq 160^\circ$, where $-1000\, {\rm km\,s^{-1}} \la V_r \la
1100\, {\rm km\,s^{-1}}$. The actual range of $V_r$-values leading to
physically acceptable solutions is therefore somewhat smaller than the
range considered for the Galactic motion calculations presented in
Section~\ref{motion}. For case~B disk crossings, the $V_r$-range is
widest when $\Omega \simeq 80^\circ$ or $\Omega \simeq 260^\circ$,
where $0\, {\rm km\,s^{-1}} \la V_r \la 550\, {\rm
km\,s^{-1}}$. Hence, for case~B, the physical constraints on the
progenitor of PSR\,J0737-3039 do not put any further constraints on
the radial velocity besides those already found from the system's
kinematic history (see Fig.~\ref{tauk1}).

In Paper~I, the mass of pulsar~B's helium star progenitor was
constrained to be between $2.1\,M_\odot$ and $4.7\,M_\odot$. These
limits correspond to the lowest mass for which a helium star is
expected to form a NS and to the highest mass for which mass transfer
from pulsar~B's helium-star progenitor to pulsar~A is expected to be
dynamically stable. Despite the dependence of the analysis presented
here on the two unknown parameters $\Omega$ and $V_r$, the range of
possible $M_0$-values is much more tightly constrained than in Paper~I
for many combinations of $\Omega$ and $V_r$. This particularly applies
to $(\Omega, V_r)$ pairs for which the kinematic age is close to the
age of 100\,Myr adopted there (see, e.g., the case~A solutions
associated with $\Omega=0^\circ$ and $80\,{\rm Myr} \le \tau_{\rm kin}
\le 100\,{\rm Myr}$ in Fig.~\ref{pro1}). The range of available helium
star masses $M_0$ furthermore tends to be most constrained for small
values of the radial velocity.

For a given post-SN orbital separation $A$ and eccentricity $e$, the
possible values of the pre-SN orbital separation $A_0$ range from
$A(1-e)$ to $A(1+e)$ (Flannery \& van den Heuvel 1975). In Paper~I, we
used the values of $A$ and $e$ obtained by integrating the equations
governing the orbital evolution due to gravitational radiation
backwards in time up to an age of 100\,Myr to constrain $A_0$ to be
between $1.36\,R_\odot$ and $1.72\,R_\odot$. These small pre-SN
orbital separations imply that the helium star progenitor of pulsar~B
is very likely overflowing its Roche lobe (see also Dewi \& van den
Heuvel 2004). Here we take into account that the age of
PSR\,J0737-3039 may be shorter than 100\,Myr. Consequently, the lower
limit on $A_0$ may actually be even smaller than $1.36\,R_\odot$ since
less orbital evolution may have taken place. An absolute lower limit
on $A_0$ is given by $A_{\rm cur}(1-e_{\rm cur})=1.15\,R_\odot$,
corresponding to the lower limit associated with the post-SN orbital
parameters for the minimum age $\tau_{\rm kin}=0$\,Myr. Since the
increase in $A$ and $e$ due to the orbital evolution backwards in time
is largest when $\tau_{\rm kin} = \tau_b = 100$\,Myr, the value of
$1.72\,R_\odot$ provides an absolute upper limit on $A_0$. Hence, for
the pre-SN helium star mass range quoted above, the conclusion reached
in Paper~I that the immediate progenitor of pulsar~B is most likely
overflowing its Roche lobe remains valid.  The range of possible
pre-SN orbital separations associated with given values of $\Omega$
and $V_r$ is furthermore {\em most restricted for the youngest
kinematic ages} found: the width of the allowed interval is
approximately $0.25\,R_\odot$ when $\tau_{\rm kin} \la 20$\,Myr and
$0.36\,R_\odot$ when $\tau_{\rm kin} \simeq 100$\,Myr.

In Paper I we derived a constrained range for the kick-velocity
magnitude between 60\,km\,s$^{-1}$ and 1560\,km\,s$^{-1}$. The lower
limit arises from the requirement that a kick counteracts the SN mass
loss and leaves the post-SN system in a tight orbit with a
sufficiently small eccentricity; the upper limit arises from the
requirement that the binary remains bound. Here we find that, for a
given radial velocity, the magnitude of the kick velocity is much
better constrained: typically to an interval that is about $\simeq
500$\,km\,s$^{-1}$ wide. If the parameter space is restricted to
kinematic ages close to 100\,Myr, as in Paper~I, the range of allowed
kick-velocity magnitudes furthermore narrows from $60\,{\rm
km\,s^{-1}} \la V_r \la 1560\,{\rm km\,s^{-1}}$ to $60\,{\rm
km\,s^{-1}} \la V_r \la 710\,{\rm km\,s^{-1}}$ for case~A disk
crossings and to $60\,{\rm km\,s^{-1}} \la V_r \la 850\,{\rm
km\,s^{-1}}$ for case~B disk crossings. In addition, the kick velocity
is smallest when $V_r$ is small and increases with increasing absolute
values of $V_r$ (see also Wex et al.\ 2000). Since the post-SN
peculiar velocity $V_{\rm pec}$ generally increases with increasing
absolute values of $V_r$ (see Fig.~\ref{vcm1}), this behavior is
consistent with Eq.~(\ref{vpec}). The highest kick velocities
furthermore tend to be associated with the youngest ages.

For case~A, we find an absolute lower limit on the kick velocity
imparted to pulsar~B of $\simeq 60$\,km\,s$^{-1}$ and an absolute
upper limit of $\simeq 1660$\,km\,s$^{-1}$; while for case~B the
absolute lower and upper limits are $\simeq 60$\,km\,s$^{-1}$ and
$\simeq 1390$\,km\,s$^{-1}$.  The lower limits are in agreement with
the lower limit derived in Paper~I and with the lower limit derived by
Dewi \& van den Heuvel (2004). This conformity arises because of the
moderate post-SN peculiar velocities associated with the Galactic disk
crossings found for $|V_r| \la 100$\,km\,s$^{-1}$ (see
Fig.~\ref{vcm1}), which impose no additional constraints on the system
in comparison to those already imposed in Paper~I. Higher absolute
values of the radial velocity would impose additional constraints and
shift the minimum kick velocity to higher values.  Ransom et
al. (2004) furthermore derived a minimum kick velocity of
105\,km\,s$^{-1}$ by assuming that the scintillation velocity
component $V^\perp$ of the {\it current} systemic velocity is entirely
due to the kick imparted to the second-born NS (i.e., that the post-SN
peculiar velocity is unaffected by the motion in the Galaxy). If we
adopt a value of $V^\perp=90$\,km\,s$^{-1}$, corresponding to the
lowest post-SN peculiar velocities found from the Galactic motion
calculations (see Fig.~\ref{vcm1}), in conjunction with an upper limit
of $4.7\,M_\odot$ on the mass of pulsar~B's helium star progenitor,
the procedure adopted by Ransom at al. (2004) yields a minimum kick
velocity of 75\,km\,s$^{-1}$. This value is in close agreement with
the minimum kick velocity derived above.

The upper limits derived for the magnitude of the kick velocity are
determined by the condition that the binary must remain bound after
the SN explosion. The condition is expressed by the inequality
\begin{equation}
V_k \le \left[
  {{G \left( M_A + M_0 \right)} \over A_0} \right]^{1/2}
  \! \left[ 1 + \left( \! 2\, {{M_A + M_B} \over
  {M_A + M_0}} \right)^{1/2} \right]  \label{vkmax}
\end{equation}
(e.g. Brandt \& Podsiadlowski 1995, Kalogera \& Lorimer 2000). For a
given set of NS masses $M_A$ and $M_B$,the upper limit increases with
increasing values of $M_0$ and with decreasing values of $A_0$. Since
the absolute lower limit on $A_0$ derived here is somewhat lower than
in Paper~I, the absolute upper limit of $\simeq 1660$\,km\,s$^{-1}$
(case~A) on $V_k$ is somewhat larger.

\begin{figure}
\resizebox{\hsize}{!}{\includegraphics{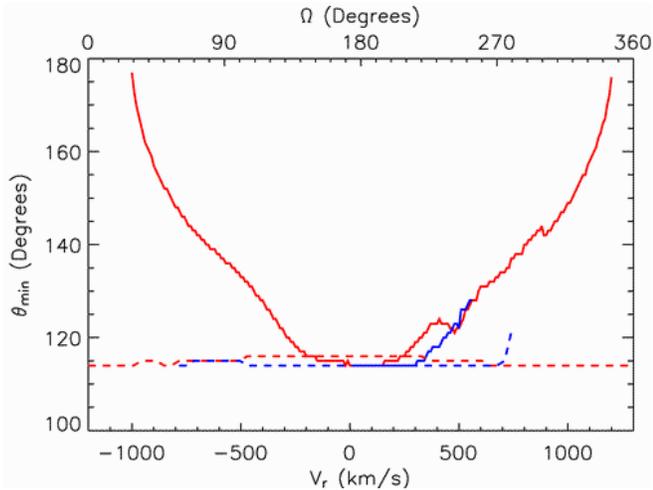}}
\caption{Minimum value of the polar angle $\theta$ between the pre-SN
  orbital velocity and the kick velocity imparted to PSR\,J0737-3039B
  at its formation as a function of $V_r$ (solid lines, bottom axis)
  and $\Omega$ (dashed lines, top axis). Case~A and~B are represented
  by black and light gray lines (red and blue in the electronic
  edition), respectively.} 
\label{pro3b}
\end{figure}

The constraints on the direction of the kick velocity imparted to
pulsar~B at birth are illustrated in Fig.~\ref{pro3b}, where we show
the minimum value $\theta_{min}$ of the angle $\theta$ between the
kick velocity and the helium star's pre-SN orbital velocity for which
all the imposed constraints may be satisfied\footnote{The maximum
possible value of $\theta$ is usually close to $180^\circ$ so that no
additional constraints arise from considering $\theta_{max}$.}. The
solid lines show the values of $\theta_{min}$ as a function of $V_r$,
while the dashed lines show the values of $\theta_{min}$ as a function
of $\Omega$. Case~A and case~B are distinguished by black and light
gray lines (red and blue in the electronic edition), respectively. The
values of $\theta_{min}$ are fairly 
insensitive to the value of $\Omega$. The dependency on $V_r$, on the
other hand, is much stronger: $\theta_{min}$ reaches a minimum value
of $\simeq 115^\circ$ for $|V_r| \la 200$\,km\,s$^{-1}$ and then
increases with increasing absolute values of $V_r$. The tendency of
the kicks to be directed opposite to the orbital motion therefore
becomes stronger with increasing absolute values of the radial
velocity (and thus with increasing values of the kick velocity, see
Fig.~\ref{pro3}).

\begin{figure}
\resizebox{\hsize}{!}{\includegraphics{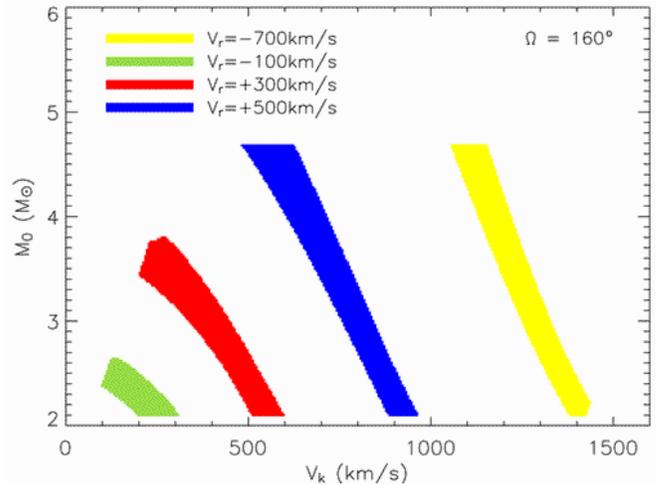}}
\caption{Correlations between the mass $M_0$ of PSR\,J0737-3039B's
  pre-SN helium star progenitor and the kick velocity $V_k$ imparted
  to its center-of-mass at the time of its formation, for case A and
  $\Omega=160^\circ$. The different gray shades (colors in the
  electronic edition) correspond to different values of $V_r$.} 
\label{m0vk1}
\end{figure}

It is also interesting to examine how the constraints on the
helium-star mass $M_{0}$ and kick magnitude $V_{k}$ correlate. For
both cases A and B, all pairs of $\Omega$ and $V_r$ yield results
qualitatively similar to the ones illustrated in Fig.~\ref{m0vk1}. It
is evident that, for a given radial velocity, higher masses correlate
with lower kicks. This is consistent with our expectations, since
stronger SN mass loss requires a smaller kick to achieve the same
effect (satisfying all of the constraints). Also as expected, as the
absolute value of $V_{r}$ increases, the constrained band shifts to
higher and higher kick magnitudes. Finally, we note that these
correlations are derived based on the constraints relevant to the
formation of PSR\,J0737-3039 and do not necessarily reflect a
correlation inherent to the kick mechanism. However, a slope for this
unknown correlation (because of the physical origin of the kick)
similar to the one shown in Fig.~\ref{m0vk1} would require minimal
fine-tuning of parameters for the formation of PSR\,J0737-3039, and
hence would be favored statistically.

\subsection{Isotropic Kick Probability Distribution}
\label{pdfvk}

\begin{figure*}
\resizebox{\hsize}{!}{\includegraphics{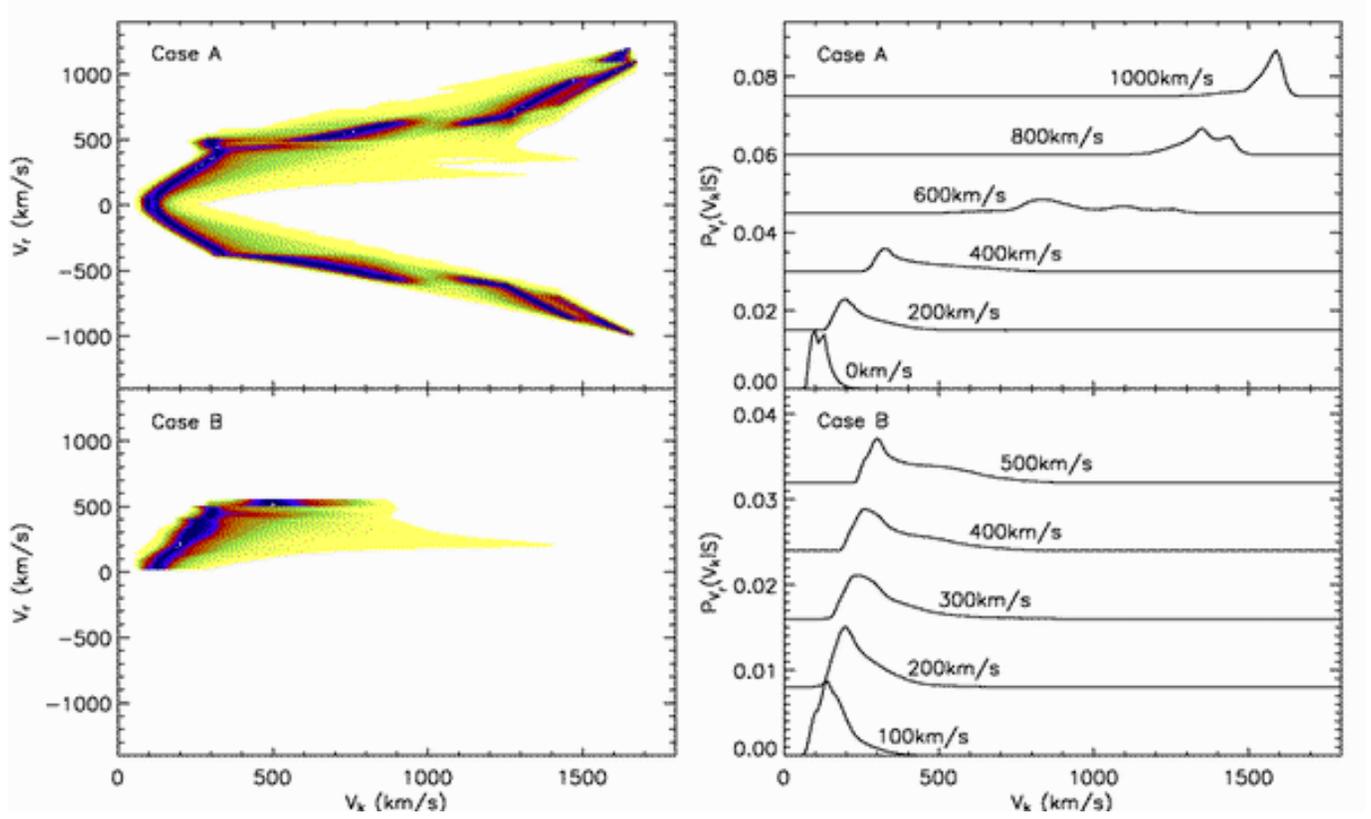}}
\caption{Probability distribution functions for the magnitude of the
kick velocity imparted to PSR\,J0737-3039B at the time of its
formation, for case~A (top panels) and~B (bottom panels). The
left-hand panels show the entire set of PDFs associated with all
admissible $V_r$-values by means of a linear color scale which varies
from light to dark gray with increasing PDF-values (in the electronic
edition increasing PDF values are indicated by yellow, green, orange,
red, and blue), while the right-hand panels show the PDFs associated
with some specific $V_r$-values. For clarity, the curves in the
right-hand panels are offset from each other by an arbitrary amount.}
\label{vk1}
\end{figure*}

In Paper~I, we derived a probability distribution for the magnitude of
the kick velocity imparted to pulsar~B, under the assumption that all
kick directions are equally probable. Here, we adopt the same
assumption and derive such probability distributions for each
combination of $\Omega$ and $V_r$ leading to plausible progenitors of
PSR\,J0737-3039. As mentioned before, for either case A or B, each
pair ($\Omega,V_{r}$) determines the kinematic age and hence the
post-SN orbital parameters $(A,e)$. Furthermore, for a given
kick-velocity magnitude $V_k$, a pair of kick-orientation angles
($\theta,\phi$) corresponds to a finite set of pre-SN properties
($M_{0},A_{0}$). Therefore, for each pair of ($\Omega,V_{r}$) and for a
given value of $V_{k}$, we can calculate the likelihood of a pair
($\theta,\phi$) that satisfies the $M_{0},A_{0}$ constraints as
\begin{eqnarray}
\Lambda_{V_r,\Omega}(\theta,\phi|V_k) 
  & \equiv & \Lambda(\theta,\phi|V_k,V_r,\Omega) \nonumber \\ 
  & = & {{j} \over {2\pi}} {{\sin \theta} \over { 2 }}, 
  {\rm ~if~} M_{0}(\theta,\phi) {\rm ~and~} A_{0}(\theta,\phi) \nonumber \\ 
  & & ~~~~~~~~~~~~~{\rm satisfy~the~constraints} \nonumber \\
  & = & 0, {\rm ~~~~~~~~~~if~not},
\label{Ba2}
\end{eqnarray}
where the polar angle $\theta$ is determined by Eq.~(\ref{eq1}), and
$j$ denotes the number of allowed ($M_{0},A_{0}$) solutions
associated with ($\theta,\phi$) for the considered $V_k$.

Using Bayes' theorem, the likelihood in Eq.~(\ref{Ba2}) is transformed
into a probability that, for a given pre-SN orbital separation $A_0$
and helium-star mass $M_0$, the kick-velocity imparted to pulsar~B had
a magnitude $V_k$:
\begin{eqnarray}
\lefteqn{P_{V_r,\Omega}(V_k|\theta,\phi) }  \nonumber \\
 & & = {{P_{V_r,\Omega}(V_k)\, 
  \Lambda_{V_r,\Omega}(\theta,\phi|V_k)} \over {\int_0^\infty 
  \Lambda_{V_r,\Omega}(\theta,\phi|V_k^\prime)\,
  P_{V_r,\Omega}(V_k^\prime)\,dV_k^\prime}}
  \label{Ba1}
\end{eqnarray}
(e.g. Bevington \& Robinson 2002).  Here $P_{V_r,\Omega}(V_k)$ is the
probability that, for a given $V_r$ and $\Omega$, the kick velocity
had a magnitude $V_k$ when no particular set of pre- and post-SN
binary parameters is imposed. Since, in this case, there is no reason
to favor any kick-velocity magnitude over another, we set
$P_{V_r,\Omega}(V_k)=1$.
  
The total probability that the kick velocity imparted to pulsar~B had
a magnitude $V_k$ is obtained by deriving a distribution
function $P_{V_r,\Omega}(V_k|\theta,\phi)$ for each admissible 
($A_0,M_0$) pair and performing the integration
\begin{equation}
P_{V_r,\Omega}(V_k|S) = \int_S P_{V_r,\Omega}(V_k|\theta,\phi)\, 
  d\theta\, d\phi.  \label{Ba3}
\end{equation}
Here, $S$ denotes the region of $(\theta,\phi)$-values corresponding 
to the entire region of admissible $A_0$- and $M_0$- values in the
constrained $(A_0,M_0)$-parameter space.  

To obtain a set of kick-velocity distributions associated with
different values of $V_r$, finally, we assume the
longitude of the ascending node to be uniformly distributed in the
plane perpendicular to the line-of-sight, and integrate the
distribution functions $P_{V_r,\Omega}(V_k|S)$ over all possible
values of $\Omega$:
\begin{equation}
P_{V_r}(V_k|S) = {1 \over {2\pi}}
  \int_0^{2\pi} P_{V_r,\Omega}(V_k|S)\, d\Omega.  \label{Ba4}
\end{equation}

The probability distribution functions $P_{V_r}(V_k|S)$ are displayed
in the left-hand panels of Fig.~\ref{vk1} by means of a linear color
scale varying from light to dark gray with increasing PDF-values (in
the electronic edition increasing PDF values are indicated by yellow,
green, orange, red, and blue). We stress that the figure depicts a set
of individual one-dimensional kick-velocity distributions associated
with different $V_r$-values, and not a single two-dimensional
distribution function of $V_k$ and $V_r$. The right-hand panels of the
figure show the probability distribution functions (PDFs) associated
with some specific values of $V_r$.

\begin{figure*}
\resizebox{\hsize}{!}{\includegraphics{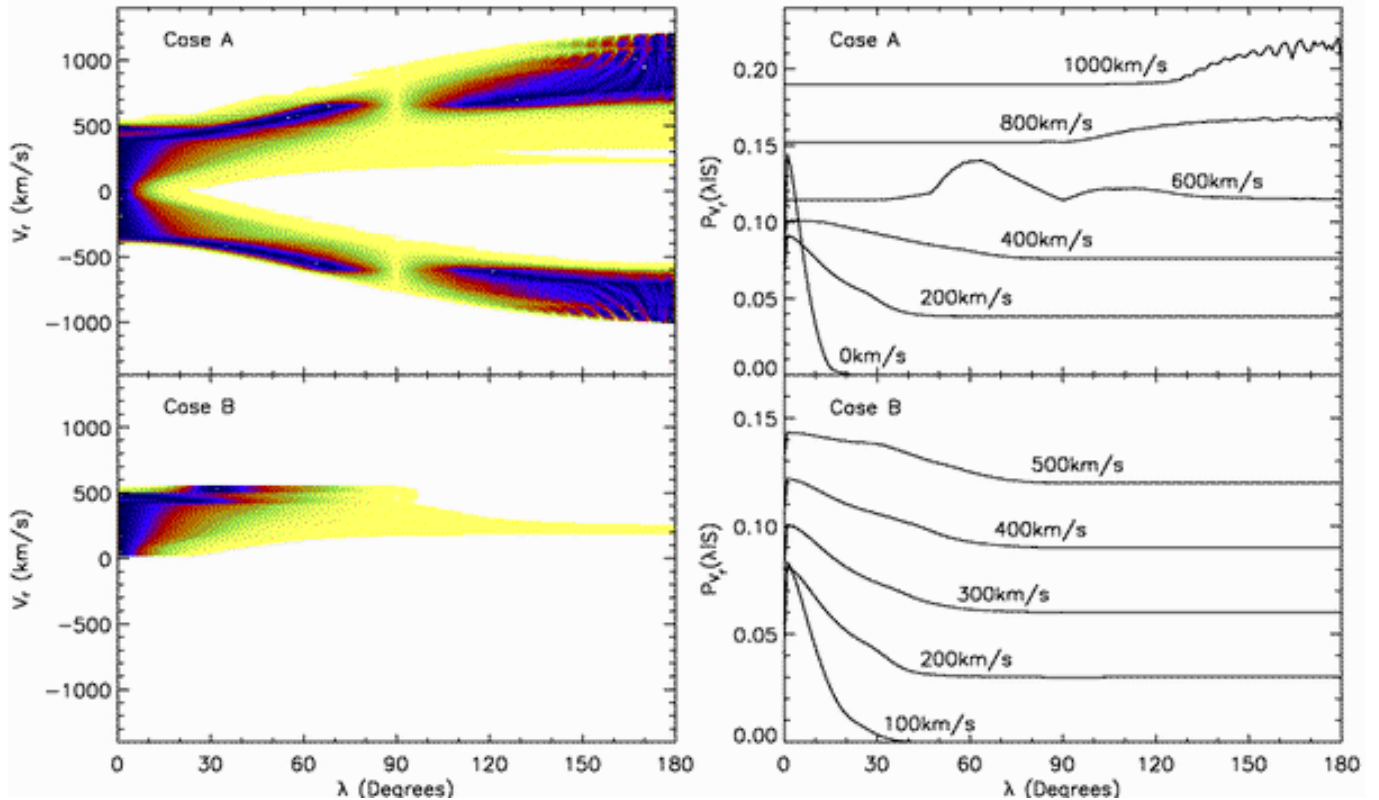}}
\caption{Probability distribution functions for the misalignment angle
$\lambda$ between PSR\,J0737-3039A's spin axis and the post-SN orbital
angular momentum axis, for case~A and~B. The left-hand panels show the
entire set of PDFs associated with all admissible $V_r$-values by
means of a linear color scale which varies from light to dark gray
with increasing PDF-values (in the electronic edition increasing PDF
values are indicated by yellow, green, orange, red, and blue), while
the right-hand panels show the PDFs associated with some specific
$V_r$-values. For clarity, the curves in the right-hand panels are
offset from each other by an arbitrary amount.}
\label{tilt1}
\end{figure*}

For case~A disk crossings, the majority of the distributions peak at a
{\it single} value of $V_k$, which increases with increasing absolute
values of $V_r$ (cf. Fig.~\ref{pro3}). This single-peaked behaviour is
in contrast to the behaviour found in Paper~I, where the kick-velocity
distribution showed a primary peak around 150\,km\,s$^{-1}$, and a
smaller secondary peak at $\simeq $1200--1300\,km\,s$^{-1}$. The
disappearance of the double-peaked structure is associated with the
additional constraints imposed by the post-SN peculiar velocities
derived from the Galactic motion calculations: since kick velocities
of the order of 150\,km\,s$^{-1}$ give rise to very different post-SN
peculiar velocities than kick velocities of the order of
1200--1300\,km\,s$^{-1}$, the extra condition effectively eliminates
one of the two peaks in favor of the other one. When $|V_r| \la
400$\,km\,s$^{-1}$, the peak in the distribution functions occurs at
kick velocities of $\simeq $100--300\,km\,s$^{-1}$, while for $|V_r|
\ga 1000$\,km\,s$^{-1}$ it occurs at kick velocities in excess of
1500\,km\,s$^{-1}$. For intermediate values of $|V_r|$ ($\simeq
600$--800\,km\,s$^{-1}$), a hint of a double-peaked structure is still
present, although the two peaks are much more evenly matched in height
and occur much closer to each other than those found in paper~I.

For case~B disk crossings, the distribution functions always show a
single-peaked behaviour. The most probable kick velocity again
increases with increasing absolute values of $V_r$ and, for $V_r \la
500\,{\rm km\,s^{-1}}$, ranges from approximately 150\,km\,s$^{-1}$ to
300\,km\,s$^{-1}$. For larger absolute values of the radial velocity,
the most probable kick velocity is about $\simeq$500\,km\,s$^{-1}$,
although only few case~B disk crossings occur for these high radial
velocities.

Since the probability of a given $V_k$ depends on the constraints
derived for $A_0$ and $M_0$, we tested the robustness of the
distribution functions to some of the assumptions adopted in the
previous sections. In particular, we varied the constraints on $A_0$
by considering both a lower upper limit of 50\,Myr and a higher upper
limit of 150\,Myr on the age of the system\footnote{Recall that the
age affects the range of possible $A_0$-values through the constraint
$A(1-e) \le A_0 \le A(1+e)$, where $A$ and $e$ are the post-SN orbital
parameters.}. We find that neither the lower nor the higher age yields
significant changes in the distribution functions, so that our results
are insensitive to reasonable changes in the constraints derived for
$A_0$. In addition, the lower limit of $2.1\,M_\odot$ on the mass
$M_0$ of pulsar~B's helium star progenitor was determined by the
lowest mass for which a {\it non-Roche-lobe-filling} helium star is
expected to form a NS. However, since the pre-SN system is most likely
undergoing a mass-transfer phase which was initiated after the
cessation of core helium burning, the lower limit of $2.1\,M_\odot$
should actually be applied to the mass of the helium star at the onset
of Roche-lobe overflow. The mass $M_0$ at the time of the second SN
explosion may then actually be somewhat lower than $2.1\,M_\odot$. We
have therefore reconsidered the derivation of the distribution
functions for a lower limit on $M_0$ of $1.4\,M_\odot$ and did not
find any significant changes in the PDFs. The reason for this is
twofold. Firstly, the increase in the admissible $(A_0,M_0)$-parameter
space resulting from a decrease of the lower limit on $M_0$ is small
in comparison to the total size of the parameter space. A decrease in
the minimum kick velocity associated with a decrease in the minimum
helium star mass (cf. Paper~I) therefore has a very low weight in the
construction of the PDFs. Secondly, for $M_0 \la 1.8\,M_\odot$ the
minimum kick velocity is no longer determined by the lower limit on
$M_0$ but by the constraints on the post-SN peculiar velocity $V_{\rm
pec}$. This is readily seen from Eq.~(\ref{vpec}) when one uses the
inequalities $V_{\rm pec} \ga 90\,{\rm km\,s^{-1}}$, $A_0 \la
1.72\,R_\odot$, and $A \la 1.54\,R_\odot$. Finally, the upper limit on
$M_0$ was varied by considering different critical mass ratios
separating dynamically stable from dynamically unstable Roche-lobe
overflow. The behavior of the PDFs again did not show any significant
changes.

\subsection{Tilt Probability Distributions} 
\label{pdftilt}

According to our current understanding of the formation of DNS
binaries, the mass-transfer phase responsible for spinning up pulsar~A
to millisecond periods is also expected to align pulsar~A's spin-axis
with the pre-SN orbital angular momentum axis. Depending on its
magnitude and orientation, the SN kick imparted to pulsar~B at the
time of its formation may tilt the post-SN orbit with respect to the
pre-SN orbital plane, thus causing a misalignment between pulsar~A's
spin axis and the post-SN orbital angular momentum axis. The tilt
angle $\lambda$ is related to the pre- and post-SN binary parameters
and to the magnitude and direction of the kick velocity as 
\begin{eqnarray}
\lefteqn{ \cos \lambda = \left[ 
  {A \over A_0}\, {{M_A + M_B} \over {M_A + M_0}}
  \left( 1-e^2 \right) \right]^{-1/2} } \nonumber \\
 & & \times \left( {V_k \over V_0}\, \cos \theta + 1 \right)
  \hspace{3.0cm} \label{tilt}
\end{eqnarray}
(Kalogera 2000). The dependence of the tilt angle on the direction of
the kick velocity is thus entirely through the polar angle $\theta$
with no contribution whatsoever from the azimuthal angle $\phi$.

In order to derive a set of probability distribution functions
$P_{V_r}(\lambda|S)$ for the tilt angle $\lambda$, we use
Eq.~(\ref{tilt}) to eliminate $V_k$ from Eqs.~(\ref{eq1})
and~(\ref{eq2}). After the elimination, the derivation of the
distribution functions is similar to the derivation of the the
kick-velocity distributions $P_{V_r}(V_k|S)$ outlined in
Section~\ref{pdfvk}. The resulting PDFs are presented in
Fig.~\ref{tilt1} in a similar fashion as the kick-velocity
distributions shown in Fig.~\ref{vk1}.

For case~A disk crossings, the tilt-angle distributions generally show
a single peak which becomes wider and moves to larger tilt angles with
increasing absolute values of $V_r$. For $|V_r| \la
400$\,km\,s$^{-1}$, the peak occurs at tilt angles $\lambda \la
30^\circ$; while for $|V_r| \ga 1000$\,km\,s$^{-1}$ it occurs at
$\lambda \ga 135^\circ$ although it becomes much less pronounced. For
intermediate $V_r$-values ($\simeq 600$\,km\,s$^{-1}$), a hint of a
double-peaked structure is present. Tilt angles close to $\lambda
\approx 90^\circ$ are furthermore strongly disfavored regardless of
the value of the radial velocity because they require greatly
fine-tuned kick magnitudes and directions satisfying $V_k\, \cos
\theta \approx -V_0$ (see Eq.~\ref{tilt}).  Case~B disk crossings, on
the other hand, yield distributions which typically peak at tilt
angles $\lambda \la 30^\circ$. The peak is most pronounced when $V_r
\la 400$\,km\,s$^{-1}$ and flattens somewhat for $V_r \ga
400$\,km\,s$^{-1}$. In the latter case, the distributions still favor
tilt angles below $70^\circ$. As for the kick-velocity distributions,
these results are very insensitive to the assumptions adopted in the
derivation of the $A_0$- and $M_0$-constraints, for both case~A and
case~B. The overall behavior of the tilt-probability distributions is
furthermore in excellent agreement with the analytical spin-tilt
distributions in coalescing binaries with two compact objects derived
by Kalogera (2000).

We conclude this section by noting that the geometrical model
constructed by Jenet \& Ransom (2004) to explain the flux variations
of PSR\,J0737-3039B predicts tilt angles of $16^\circ \pm 10^\circ$ or
$164^\circ \pm 10^\circ$.  The PDFs presented in Fig.~\ref{tilt1}
favor the lower tilt angle when $|V_r| \la 400$\,km\,s$^{-1}$ and the
higher tilt angle when $|V_r| \ga 800$\,km\,s$^{-1}$. The alternative
solutions of $82^\circ \pm 16^\circ$ or $98^\circ \pm 16^\circ$
derived by Jenet \& Ransom (2004) are strongly disfavored by all
distribution functions. As noted by the authors, these solutions are
also incompatible with the misalignment angle between the pulsar's
spin and magnetic dipole axes derived by Demorest et al. (2004).

\subsection{Non-isotropic kicks}

The probability distribution functions for the kick-velocity magnitude
$V_k$ and the spin misalignment angle $\lambda$ presented in the
previous sections were derived under the assumption that all kick
directions are equally probable. However, in recent years a possible
alignment of the SN kick with the NS spin axis has been discussed, 
prompted mainly by the alignment of pulsar proper motion vectors with jet
axes in the Crab and the Vela pulsars (Lai, Chernoff, \& Cordes 2001;
Romani 2004).

Assuming that the NS spin axis is determined by the axis of the
progenitor helium star and that this axis was roughly aligned with the
pre-SN orbital angular momentum axis due to mass transfer, the
alignment inferred by these observations possibly indicates a
preference for kick directions perpendicular to the pre-SN orbital
plane. In order to examine the effect of such preferred kick
directions, we reconsider the derivation of the kick-velocity and
tilt-angle distributions under the assumption that the kick direction
is restricted to be within two oppositely directed cones with opening
angle $\xi_p$ and with axes parallel to the pre-SN orbital angular
momentum axis (i.e., polar kicks). For this purpose, it is convenient
to introduce the angle $\xi \in [0,\pi]$ between the kick-velocity
vector and the pre-SN orbital angular momentum vector as a measure for
the spin-kick alignment. The angle is related to the kick-direction
angles $\theta$ and $\phi$ by the equation
\begin{equation}
\cos \xi = \sin \theta\, \cos \phi,  \label{xi}
\end{equation}
so that kicks are directed within the above mentioned cones with
opening angle $\xi_p$ when 
\begin{equation}
\cos \xi_p \le |\sin \theta\, \cos \phi| \le 1  \label{cone}
\end{equation}
(Kalogera 2000). We also note that since, for a given kick-velocity
magnitude $V_k$, four solutions $(\theta,\phi)$ of Eqs.~(\ref{eq1})
and~(\ref{eq2}) exist for each set of admissible pre- and post-SN
binary parameters, there is also a two-fold degeneracy in the
solutions for $\xi$. 

\begin{figure}
\resizebox{\hsize}{!}{\includegraphics{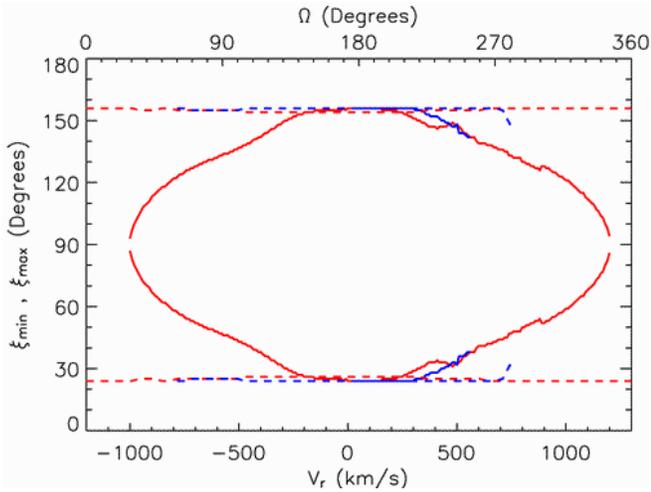}}
\caption{Limits on the polar angle $\xi$ between the pre-SN orbital
  angular momentum axis and the kick velocity imparted to
  PSR\,J0737-3039B at its formation as a function of $V_r$ (solid
  lines, bottom axis) and $\Omega$ (dashed lines, top axis). Values of
  $\xi$ smaller than $90^\circ$ are lower limits ($\xi_{min}$); values
  of $\xi$ larger than $90^\circ$ are upper limits ($\xi_{max}$).
  Case~A and~B are represented by black and light gray lines (red and
  blue in the electronic edition), respectively.}
\label{pro3c}
\end{figure}

The range of $\xi$ angles available for the formation of
PSR\,J0737-3039 is shown in Fig.~\ref{pro3c} as a function of $V_r$
and $\Omega$. Curves below $90^\circ$ correspond to lower limits,
$\xi_{min}$, and curves above $90^\circ$ to upper limits,
$\xi_{max}$. Case~A and~B are indicated by black and light gray lines
(red and blue in the electronic edition), 
respectively.  It is interesting to note that it would be impossible
to understand the formation of PSR\,J0737-3039, if the kick were very
closely aligned or anti-aligned to the pre-SN orbital angular momentum
(that could also determine the newborn NS spin). Instead a
minimum deviation from the pre-SN orbital angular momentum axis of
$25^\circ$--$30^\circ$ appears to be necessary to satisfy all the
imposed constraints.

\begin{figure*}
\resizebox{\hsize}{!}{\includegraphics{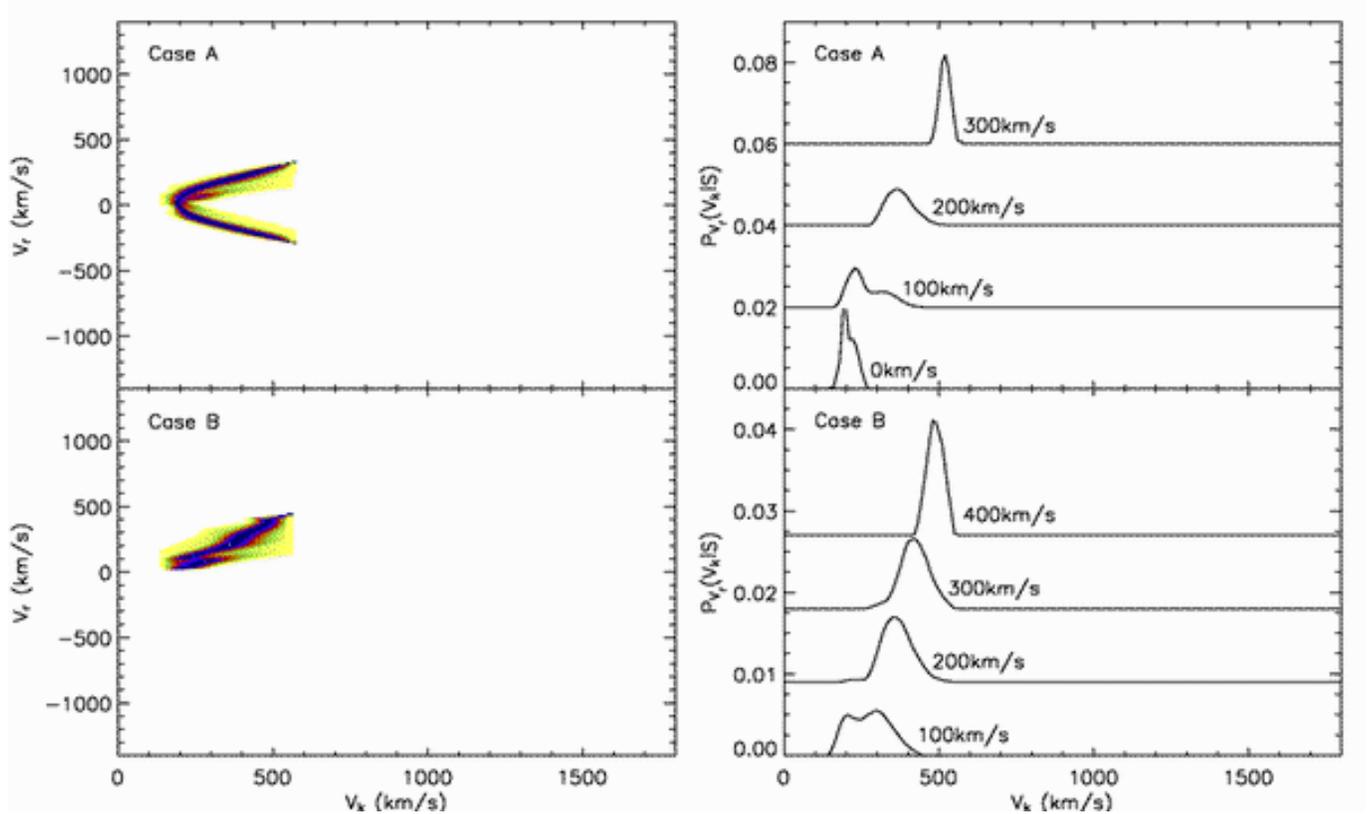}}
\caption{Probability distribution functions for the magnitude $V_k$ of
the kick velocity imparted to PSR\,J0737-3039B at the time of its
formation, for polar kicks restricted to two oppositely directed cones
with an opening angle of $\xi_p = 30^\circ$ and with axes parallel to
the pre-SN orbital angular momentum axis (cf. Fig.~\ref{vk1} for more
details).}
\label{vk1b}
\end{figure*}

\begin{figure*}
\resizebox{\hsize}{!}{\includegraphics{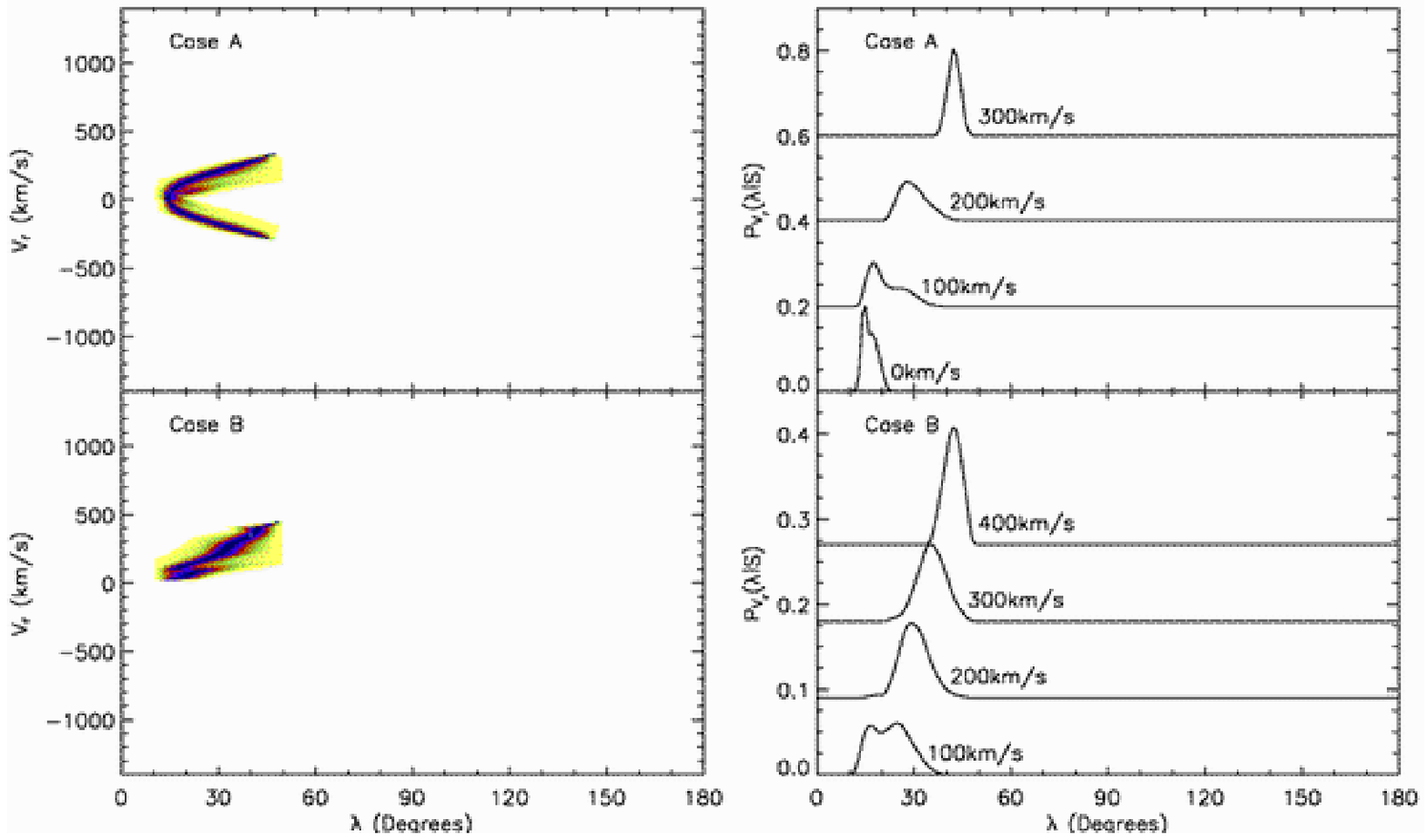}}
\caption{Probability distribution functions for the misalignment angle
$\lambda$ between PSR\,J0737-3039A's spin axis and the post-SN orbital
angular momentum axis, for polar kicks restricted to two oppositely
directed cones with an opening angle of $\xi_p = 30^\circ$ and with
axes parallel to the pre-SN orbital angular momentum axis
(cf. Fig.~\ref{tilt1} for more details).}
\label{tilt1b}
\end{figure*}

In order to illustrate the effects of polar kicks, we considered the
case where $\xi_p=30^\circ$. The reduction in the available parameter
space associated with the restriction of the kick direction yields
viable solutions for case~A only when $|V_r| \la 300$\,km\,s$^{-1}$,
and for case~B only when $0\,{\rm km\,s^{-1}} \la V_r \la 500\,{\rm
km\,s^{-1}}$. In addition, the mass of pulsar~B's helium star
progenitor becomes constrained to $M_0 \approx 2.1$--$2.5\,M_\odot$,
the pre-SN orbital separation to $A_0 \approx 1.1$--$1.5\,R_\odot$,
and the magnitude of the kick velocity to $V_k \approx
100$--600\,km\,s$^{-1}$.

In Figs.~\ref{vk1b} and~\ref{tilt1b}, we show the kick-velocity and
tilt-angle distribution functions in the case of polar kicks with
$\xi_p=30^\circ$. Both sets of distributions clearly show a much more
pronounced single peak than when the kicks are assumed to be
distributed isotropically (cf. Figs.~\ref{vk1}
and~\ref{tilt1}). Depending on the radial velocity, the most probable
kick velocity ranges from approximately 200\,km\,s$^{-1}$ to
550\,km\,s$^{-1}$, and the most probable tilt angle from $15^\circ$ to
$45^\circ$. This range of tilt angles is compatible with the tilt
angle of $16^\circ \pm 10^\circ$ predicted by Jenet \& Ransom (2004).

\section{THE RECENT EVOLUTIONARY HISTORY OF PSR\,B1534+12}
\label{sec1534}

Unlike PSR\,J0737-3039, the relativistic binary radio pulsar
PSR\,B1534+12 has an accurately measured proper motion with a known
direction in right ascension and declination (see Table~\ref{param}),
so that its kinematic history and progenitor constraints {\em depend
only on the unknown radial velocity $V_r$}. In order to derive these
constraints, we adopt the spin-down age $\tau_b=210$\,Myr as an upper
limit to the age of the system. The other physical parameters of
PSR\,B1534+12 relevant to the derivation are summarised in
Table~\ref{param}.

Following the same arguments as for PSR\,J0737-3039 in
Section~\ref{motion}, we assume that PSR\,B1534+12 was close to the
Galactic plane at the time of the second SN explosion and that its
pre-SN systemic velocity was almost entirely due to Galactic
rotation. The possible birth sites of the DNS are then obtained by
tracing the Galactic motion of the system backwards in time as a
function of the unknown radial velocity $V_r$. We find that, within
the imposed age limit of 210\,Myr, no Galactic plane crossings occur
when $V_r \la -280$\,km\,s$^{-1}$, whereas up to four crossings
may occur when $V_r \ga -280$\,km\,s$^{-1}$. Since four disk
crossings only occur for relatively few and rather fine-tuned Galactic
trajectories, we leave these cases aside and focus on the possible
birth sites associated with the first (case~A), second (case~B), and
third (case~C) Galactic plane crossings.

\begin{figure}
\resizebox{\hsize}{!}{\includegraphics{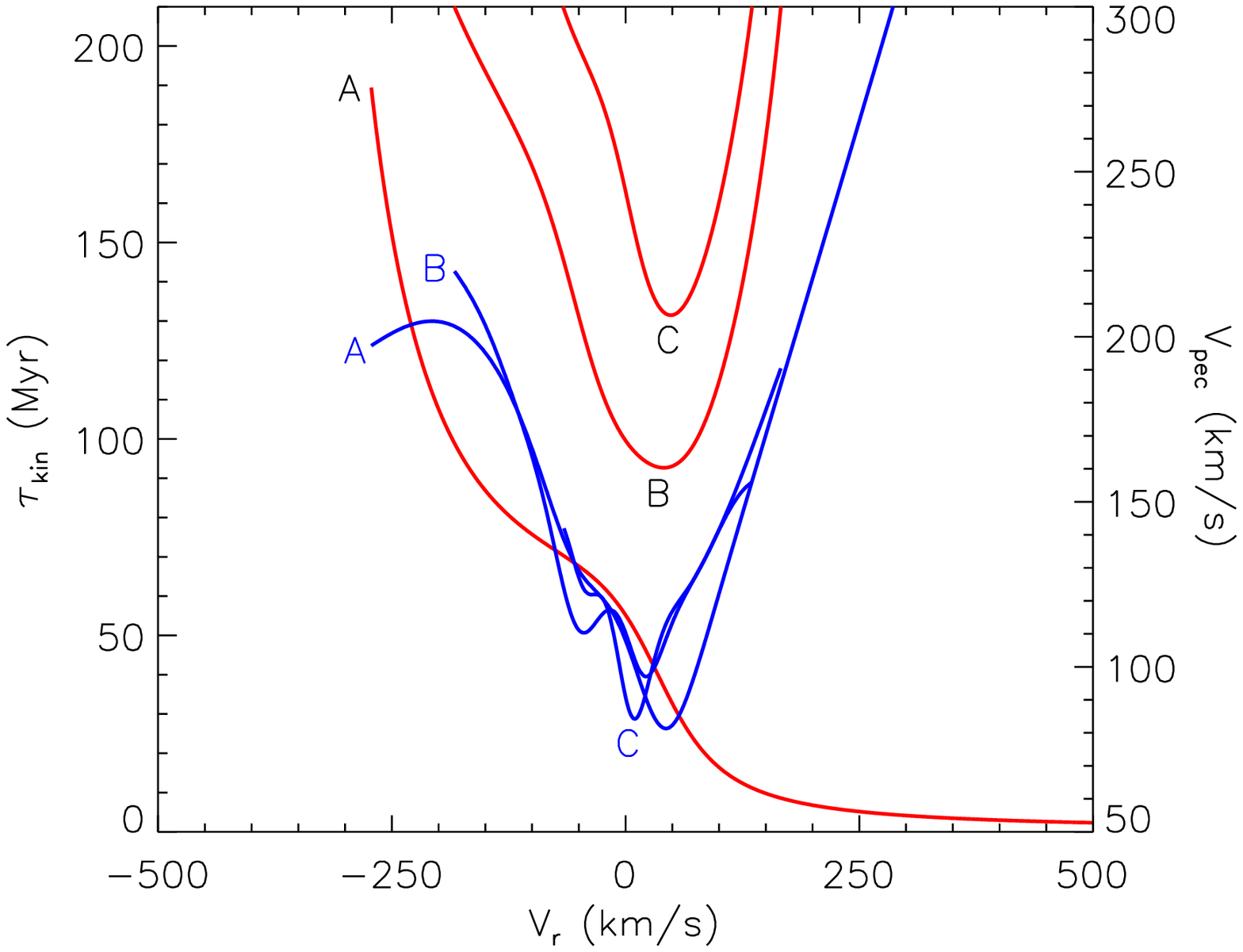}}
\caption{Variations of the kinematic age (left-hand axis)
  and post-SN peculiar velocity (right-hand axis) of
  PSR\,B1534+12 as a function of the unknown radial velocity $V_r$,
  for cases~A, B, and~C. The kinematic ages are represented by black
  solid lines (red in the electronic edition) and the peculiar
  velocities by light gray lines (blue in the electronic edition).}  
\label{tauk4}
\end{figure}

The kinematic ages and post-SN peculiar velocities associated with the
Galactic plane crossings are shown in Fig.~\ref{tauk4} as functions of
the radial velocity $V_r$. Case~A gives rise to a wide range of ages
between $\simeq 1$\,Myr and 210\,Myr, and peculiar velocities between
80\,km\,s$^{-1}$ and 1500\,km\,s$^{-1}$. Cases~B and C, on the
other hand, yield kinematic ages of {\em at least} $\simeq
90$\,Myr and $\simeq 130$\,Myr; and post-SN peculiar velocities of
100--220\,km\,s$^{-1}$ and 80--160\,km\,s$^{-1}$, respectively.

\begin{figure*}
\resizebox{\hsize}{!}{\includegraphics{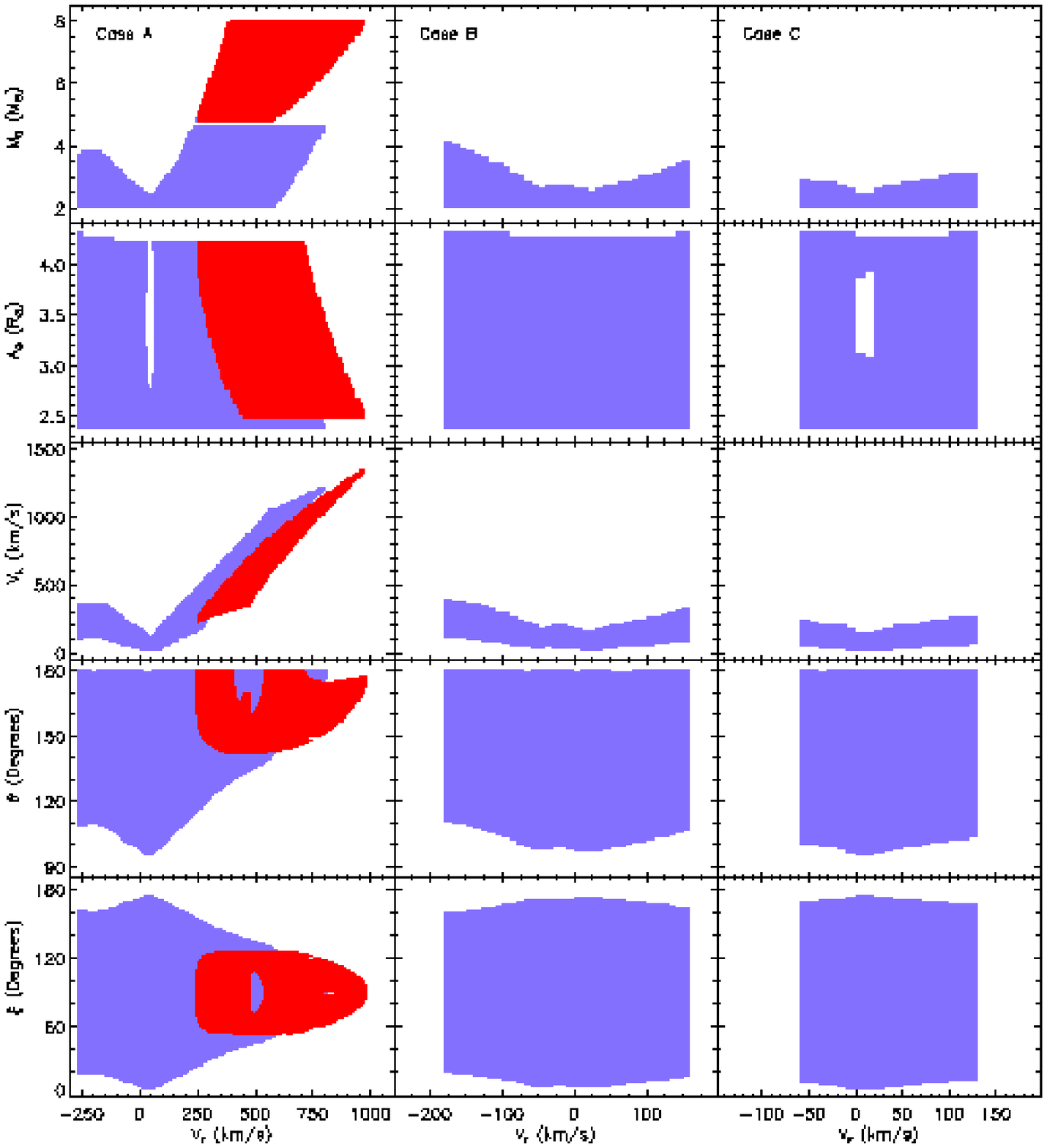}}
\caption{Limits on the pre-SN progenitor of PSR\,B1534+12 and on the
  kick velocity imparted to the last-born NS. The left-hand panels
  show the constraints for case~A, the middle panels for case~B, and
  the right-hand panels for case~C. Dark gray (red in the electronic
  edition) regions correspond to solutions for which the pre-SN binary
  is detached, while light gray (blue in the electronic edition)
  regions indicate the additional solutions associated with
  mass-transferring systems.}
\label{pro10}
\end{figure*}

The post-SN orbital separation $A$ and orbital eccentricity $e$ at the
times of Galactic disk crossings are obtained by numerical integration
backwards in time of the equations governing the evolution of the
orbit under the influence of gravitational radiation. The resulting
post-SN orbital parameters range from $A=3.28\,R_\odot$ and $e=0.274$
when $\tau_{kin}=0$\,Myr to $A=3.36\,R_\odot$ and $e=0.282$ when
$\tau_{kin}=210$\,Myr. These post-SN orbital parameters together with
the post-SN peculiar velocities impose constraints on the pre-SN
progenitor of PSR\,B1534+12. These are derived in a similar way as for
the progenitor of PSR\,J0737-3037 (except that the problem depends on
only one free parameter, $V_r$). The results of our analysis are
summarized in Fig.~\ref{pro10}.

Contrary to PSR\,J0737-3039, which was most likely in a state of mass
transfer just before the second SN explosion, PSR\,B1534+12 may have
been detached as well as semi-detached at the time the second NS was
born. In order to illustrate this, the solutions for which no mass
transfer takes place at the time of the second SN explosion are
indicated by the dark gray (red in the electronic edition) regions in
Fig.~\ref{pro10}, while the light gray (blue in the electronic
edition) regions indicates the {\it additional} solutions that become
accessible when the possibility of mass transfer is taken into
account. For the latter solutions we adopt the same assumption as
before that the system may survive the mass transfer phase and form a
DNS only if the mass ratio $M_0/M_A$ is smaller than 3.5 (e.g.,
Ivanova et al. 2003).

For case~A disk crossings, detached pre-SN binary configurations exist
only for $V_r \ga 250$\,km\,s$^{-1}$. For smaller radial
velocities, the system is always undergoing mass transfer from the
progenitor of the second-born NS to the first-born NS. The pre-SN mass
of the helium star forming the second NS is constrained within
$2.1\,M_\odot$ -- $8\,M_\odot$. The lower limit again corresponds to the
lowest mass for which a helium star is expected to form a NS instead
of a white dwarf, while the upper limit corresponds to the highest
mass for which a helium star is expected to form a NS rather than a
black hole (see, e.g., Fig.~1 in Belczynski, Kalogera, \& Bulik 2002;
and Table~16.4 in Tauris \& van den Heuvel 2004). The divide between
the dark and light gray (red and blue in the electronic edition)
regions at $4.7\,M_\odot$ corresponds to the adopted critical mass
ratio $M_0/M_A=3.5$ for dynamically stable mass transfer. The allowed
mass range of pre-SN helium star masses is most constrained for $|V_r|
\la 200$\,km\,s$^{-1}$ when $2.1\,M_\odot \la M_0 \la
4\,M_\odot$. Lower and upper limits on the pre-SN orbital separation
are given by $2.4\,R_\odot$ and $4.3\,R_\odot$.

\begin{figure*}
\resizebox{\hsize}{!}{\includegraphics{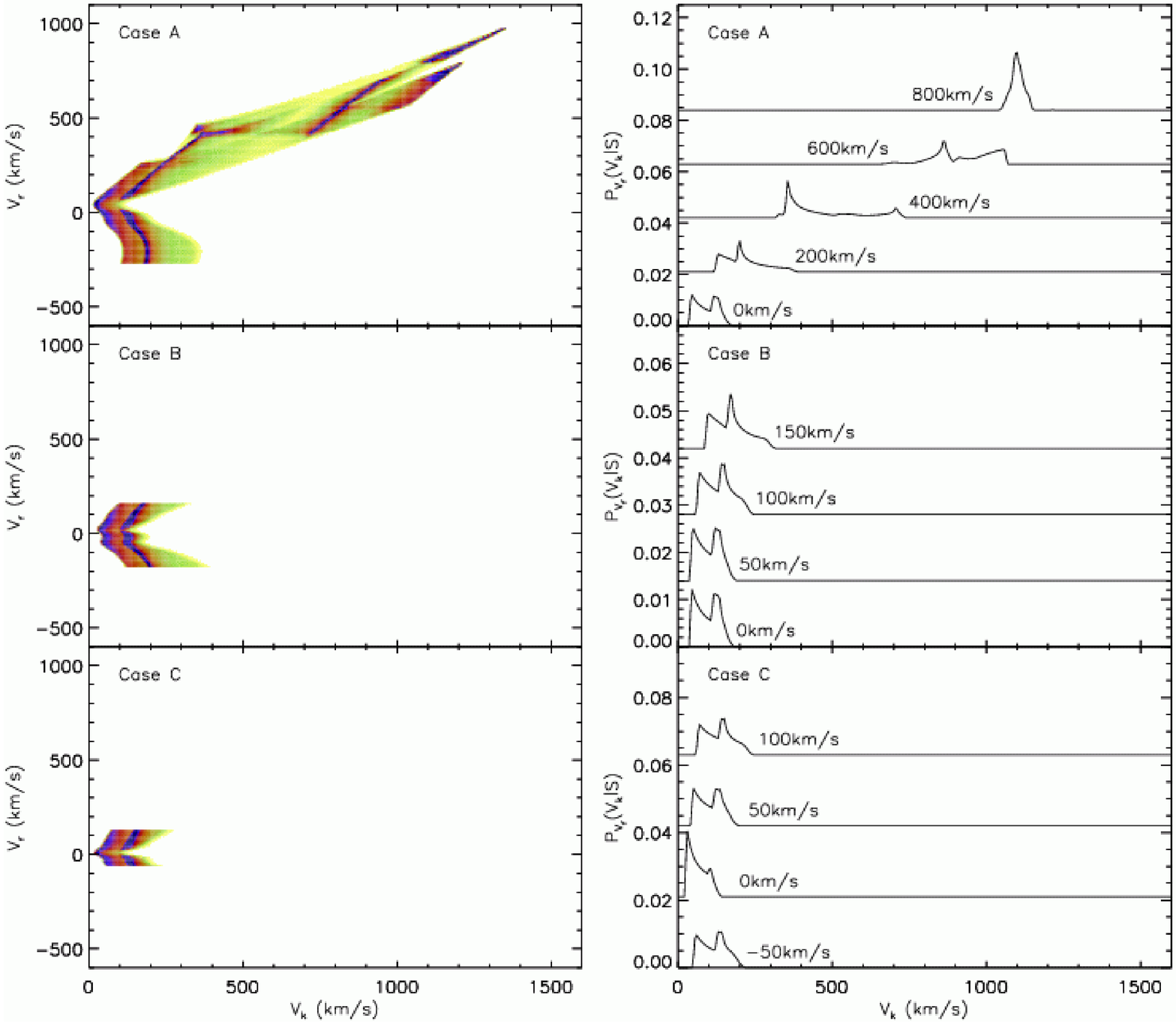}}
\caption{Probability distribution functions of the magnitude of the
kick velocity imparted to PSR\,B1534+12's companion at the time of its
formation, for cases~A, B, and~C (cf. Fig.~\ref{vk1} for more
details).} 
\label{vk4}
\end{figure*}

The behavior of the kick-velocity magnitude is similar as for
PSR\,J0737-3039: the kick velocity increases with increasing absolute
values of $V_r$ and, for a given radial velocity, it is generally
constrained to an interval that is less than $\simeq
600$\,km\,s$^{-1}$ wide.  When all possible pre-SN binary
configurations are considered, the kick velocity magnitudes range from
15\,km\,s$^{-1}$ to 1350\,km\,s$^{-1}$. When configurations for
which the helium star is overflowing its Roche lobe are excluded, the
range narrows to $230\,{\rm km\,s^{-1}} \la V_k \la 1350\,{\rm
km\,s^{-1}}$, so that the minimum kick velocity is higher than when
the possibility of Roche-lobe overflow is taken into account. The
larger minimum kick velocity is required to compensate the larger
effect of the mass lost from the system during the SN explosion: since
the minimum $M_0$ for these systems is $4.7\,M_\odot$, at least $57\%$
of the total pre-SN mass is lost from the system. The minimum and
maximum kick velocities are furthermore in good agreement with those
derived by Fryer \& Kalogera (1997) and Dewi \& Pols (2003).

The polar angle $\theta$ between the kick velocity and the helium
star's pre-SN orbital velocity is always larger than $95^\circ$,
so that the kicks tend to be directed opposite to the orbital
motion. When the solutions are restricted to those where no Roche-lobe
overflow is going on at the time of the second SN explosion, the
direction of the kick velocity is constrained to $\theta \ga
140^\circ$. These values are again in good agreement with those
derived by Dewi \& Pols (2003).  The angle $\xi$ between the kick
velocity and the pre-SN orbital angular momentum axis, on the other
hand, takes values between $5^\circ$ and $175^\circ$. When the
parameter space is restricted to the solutions where the binary is
detached just before the second SN explosion, the range of possible
$\xi$-values narrows to $55^\circ \la \xi \la 125^\circ$.

\begin{figure*}
\resizebox{\hsize}{!}{\includegraphics{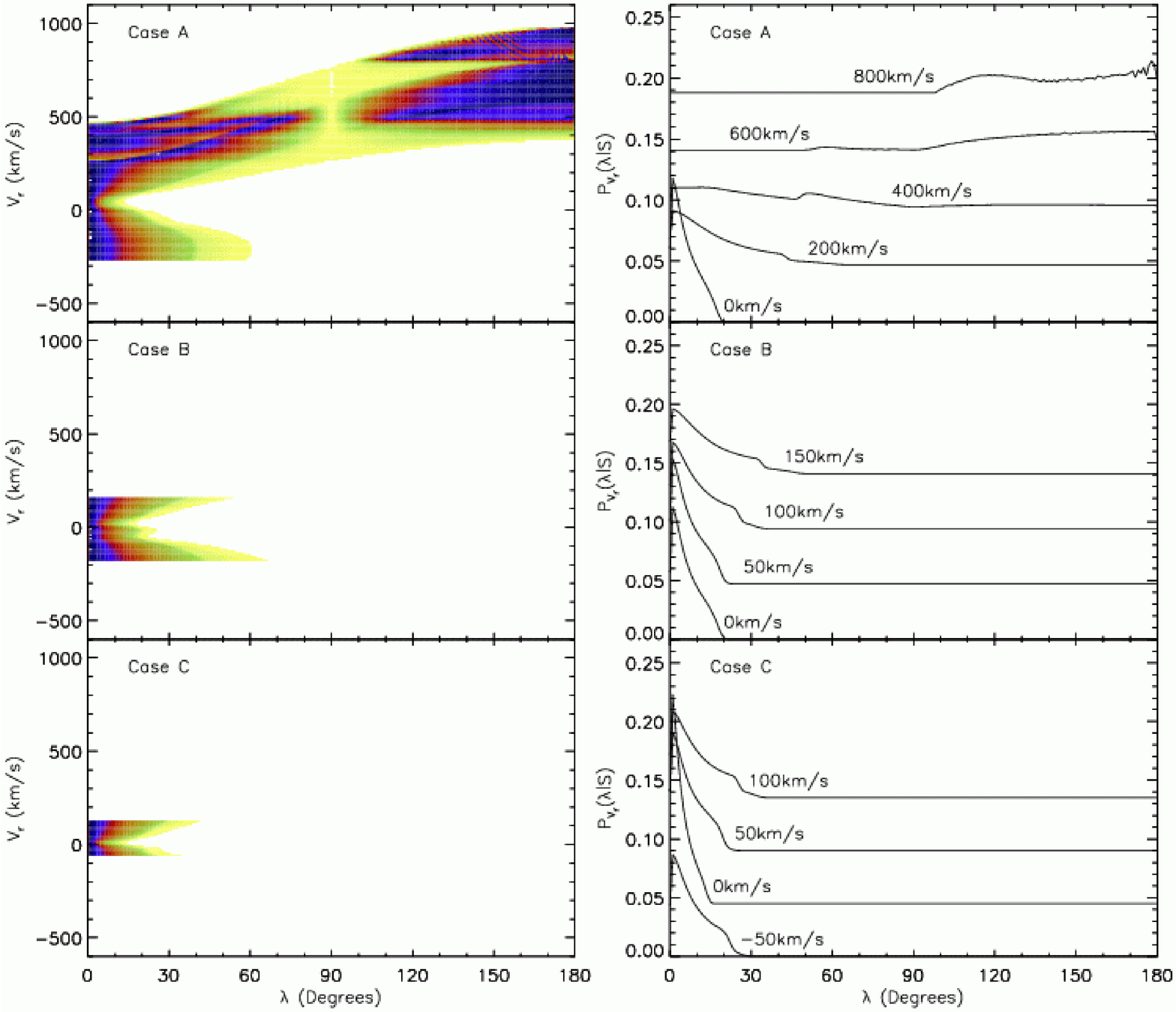}}
\caption{Probability distribution functions for the misalignment angle
$\lambda$ between PSR\,B1534+12's spin axis and the post-SN orbital
angular momentum axis, for cases~A, B, and~C (cf. Fig.~\ref{tilt1} for
more details).}
\label{tilt3}
\end{figure*}

For case~B disk crossings, all allowed pre-SN binary
configurations imply that the helium star progenitor of the
second-born NS is overflowing its Roche lobe at the time of its SN
explosion. The mass of the helium star is constrained to 
$2.1\,M_\odot \la M_0 \la 4.1\,M_\odot$, the pre-SN orbital
separation to $2.4\,R_\odot \la A_0 \la 4.3\,R_\odot$, and the
kick-velocity magnitude to $30\,{\rm km\,s^{-1}} \la V_k \la
390\,{\rm km\,s^{-1}}$. The angle between the kick direction and the
helium star's pre-SN orbital velocity is restricted $100^\circ \la
\theta \la 180^\circ$, and the angle between the kick direction and
the pre-SN orbital angular momentum axis to $10^\circ \la \xi \la
170^\circ$.

Case~C disk crossings, finally, also yield no detached solutions
for the progenitor of PSR\,B1534+12. The constraints for this case are
similar to those found for case~B, except that the mass $M_0$ is
always smaller than $\simeq 3.1\,M_\odot$ and the kick velocity
$V_k$ is always smaller than $270\,{\rm km\,s^{-1}}$.

From these constraints, it follows that, if the case for the alignment
of kicks with the spin of the NS's progenitor gets more support in the
future, the direct progenitor of PSR\,B1534+12's companion must be
overflowing its Roche lobe, regardless of which disk crossing
corresponds to the birth place. In addition, as for PSR\,J0737-3039,
the formation of PSR\,B1534+12 cannot be explained if the kick
imparted to PSR\,B1534+12's companion was {\em perfectly} aligned or
anti-aligned with the spin of its progenitor.

The constraints derived above may again be used to derive probability
distribution functions for the magnitude $V_k$ of the kick velocity
imparted to the second-born NS and for the misalignment angle
$\lambda$ between the spin-axis of the first-born NS and the post-SN
orbital angular momentum axis (cf. Sections~\ref{pdfvk}
and~\ref{pdftilt}). The resulting PDFs are presented in
Figs.~\ref{vk4} and~\ref{tilt3} as those for PSR\,J0737-3039 in
Figs.~\ref{vk1} and~\ref{tilt1}. For radial velocities of only a few
100\,km\,s$^{-1}$, the kick-velocity distributions show two closely
spaced and fairly evenly matched peaks between $V_k \simeq
50$\,km\,s$^{-1}$ and $V_k \simeq 250$\,km\,s$^{-1}$. For higher
radial velocities, relevant only to case~A, the peak(s) shift to
larger kick velocities up to a maximum of $\simeq
1350$\,km\,s$^{-1}$. The tilt-angle distributions, on the other hand,
favor misalignment angles below $30^\circ$ when $|V_r| \la
200$\,km\,s$^{-1}$ and above $100^\circ$ when $|V_r| \ga
600$\,km\,s$^{-1}$. For case~A disk crossings, tilt angles close to
$\lambda \approx 90^\circ$ are furthermore strongly disfavored
regardless of the value of the radial velocity.  For case~B disk
crossings, tilt angles with non-vanishing probabilities are always
smaller than $40^\circ$--$50^\circ$.

\section{THE RECENT EVOLUTIONARY HISTORY OF PSR\,B1913+16}
\label{sec1913}

Besides the knowledge of the proper motion direction, PSR\,B1913+16
has the advantage that the misalignment angle between the pulsar's
spin axis and the pre-SN orbital angular momentum axis has been
determined to be around $\simeq 20^\circ$, corresponding to prograde
rotation, or $\simeq 160^\circ$, corresponding to retrograde rotation
(Kramer 1998; Weisberg \& Taylor 2002). Wex et al. (2000) used this
information to derive constraints on the mass of the second-born NS's
direct progenitor, on the pre-SN orbital separation, and on the
magnitude and direction of the kick velocity imparted to the
second-born NS at birth. In agreement with the then available
numerical simulations of rapidly accreting neutron stars, the authors
assumed that mass transfer from a helium star companion would cause
the NS to collapse into a black hole and therefore excluded Roche-lobe
overflowing helium stars as viable progenitors of the second-born
NS. However, more recent calculations by Dewi et al. (2002), Ivanova
et al. (2003), and Dewi \& Pols (2003) show that NS binaries may
survive a helium-star mass-transfer phase, provided that the ratio of
the helium star's mass to the neutron star's mass is not too extreme
($M_0/M_A \le 3.5$). In this section, we therefore revise the
constraints derived by Wex et al. (2000) in the light of this new
information. The adopted physical parameters are listed in
Table~\ref{param}. The constraints on the tilt angle, for which we
consider the values $\lambda=18^\circ \pm 6^\circ$ and
$\lambda=162^\circ \pm 6^\circ$, are imposed by means of
Eq.~(\ref{tilt}).

We look for possible birth sites of PSR\,B1913+16 by tracing the
system's motion in the Galaxy backwards in time up to a maximum age of
80\,Myr\footnote{As for PSR\,J0737-3039 and PSR\,B1534+12, we use the
spin-down age $\tau_b=80$\,Myr instead of the characteristic age
$\tau_c=110$\,Myr as an upper limit for the age of the system. The
maximum amount of orbital evolution that may have taken place since
the formation of the DNS is therefore somewhat smaller than in Wex et
al. (2000).}, as a function of the unknown radial velocity $V_r$. In
agreement with Wex et al. (2000), we find that the system may have
crossed the Galactic plane up to two times. The first Galactic plane
crossing (case~A) occurs at very young kinematic ages of $\simeq
2$--4\,Myr and gives rise to peculiar velocities in excess of $\simeq
300$\,km\,s$^{-1}$. The corresponding post-SN orbital parameters are
$A \approx 2.8\, R_\odot$ and $e \approx 0.618$. The second Galactic
plane crossing (case~B), on the other hand, takes place at least
$\simeq 55$\,Myr in the past and yields peculiar velocities of $\simeq
230$--440\,km\,s$^{-1}$. The associated post-SN orbital separations
and eccentricities range from $A=3.1\,R_\odot$ and $e=0.646$ to
$A=3.2\,R_\odot$ and $e=0.658$.

\begin{figure*}
\resizebox{\hsize}{!}{\includegraphics{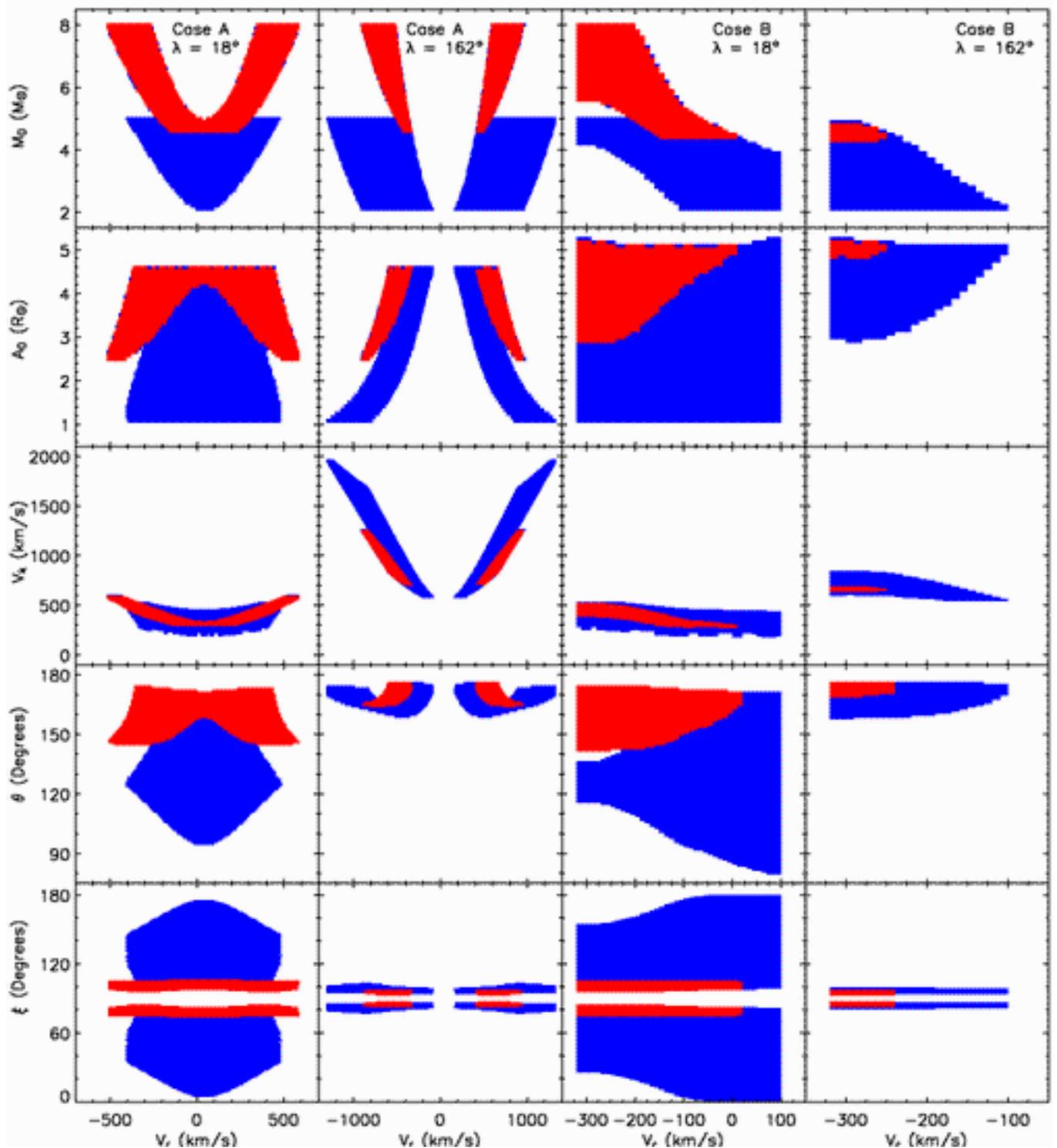}}
\caption{Limits on the pre-SN progenitor of PSR\,B1913+16 and on the
  kick velocity imparted to the last-born NS. Dark gray (red in the
  electronic edition) regions correspond to solutions for which the
  pre-SN binary is detached, while light gray (blue in the electronic
  edition) regions indicate the additional solutions associated with
  mass-transferring systems.}
\label{pro9}
\end{figure*}

The constraints on the pre-SN parameter space accessible to the
progenitor of PSR\,B1913+16 are shown in Fig.~\ref{pro9} as functions
of the unknown radial velocity $V_r$. As before, the dark gray (red in
the electronic edition) regions correspond to the solutions for which
no mass transfer takes place at the time of the second SN explosion,
while the light gray (blue in the electronic edition) regions indicate
the additional solutions found when the possibility of mass transfer
is taken into account. The constraints for detached pre-SN binary
configurations are in good agreement with the constraints derived by
Wex et al. (2000). It is clear however, that when the possibility of
mass transfer is taken into account, the available parameter space
becomes much less constrained.

For case~A disk crossings, the range of radial velocities for which
physically acceptable solutions exist is restricted to $|V_r| \la
500$--600\,km\,s$^{-1}$ when $\lambda=18^\circ$, and to $100\,{\rm
km\,s^{-1}} \la |V_r| \la 1300\,{\rm km\,s^{-1}}$ when
$\lambda=162^\circ$. The mass of the second-born NS's direct
progenitor is constrained to the interval between $2.1\,M_\odot$ and
$8\,M_\odot$, and the pre-SN orbital separation to the interval
between $1.1\,R_\odot$ and $5.3\,R_\odot$ for both considered values
of the tilt angle $\lambda$. The magnitude of the kick velocity varies
from $\simeq 190$\,km\,s$^{-1}$ to $\simeq 600$\,km\,s$^{-1}$ when
$\lambda=18^\circ$ and from $\simeq 580$\,km\,s$^{-1}$ to $\simeq
2000$\,km\,s$^{-1}$ when $\lambda=162^\circ$. When the solutions are
restricted to detached pre-SN binary configurations the limits become
$300\,{\rm km\,s^{-1}} \la V_k \la 600\,{\rm km\,s^{-1}}$ and
$700\,{\rm km\,s^{-1}} \la V_k \la 1250\,{\rm km\,s^{-1}}$ for
$\lambda=18^\circ$ and $\lambda=162^\circ$, respectively. The
direction of the kick velocity is most constrained when
$\lambda=162^\circ$: for $\lambda=18^\circ$ we have $95^\circ \la
\theta \la 175^\circ$ and $ 5^\circ \la |90^\circ-\xi| \la 85^\circ$,
while for $\lambda=162^\circ$ we find $160^\circ \la \theta \la
175^\circ$ and $5^\circ \la |90^\circ-\xi| \la 15^\circ$. When the
parameter space is restricted to detached pre-SN binary
configurations, the limits become $145^\circ \la \theta \la 175^\circ$
and $5^\circ \la |90^\circ-\xi| \la 15^\circ$ for $\lambda=18^\circ$, and
$165^\circ \la \theta \la 175^\circ$ and $|90^\circ-\xi| \approx
5^\circ$ for $\lambda=162^\circ$.

For case~B disk crossings, the mass $M_0$ of the second-born NS's
direct progenitor and the pre-SN orbital separation $A_0$ are
constrained to the ranges of values given by $2.1\,M_\odot \la M_0 \la
8\,M_\odot$ and $1.1\,R_\odot \la A_0 \la 5.3\,R_\odot$ when
$\lambda=18^\circ$, and to $2.1\,M_\odot \la M_0 \la
5\,M_\odot$ and $2.9\,R_\odot \la A_0 \la 5.3\,R_\odot$ when
$\lambda=162^\circ$. The magnitude of the kick velocity varies between
$\simeq 190$\,km\,s$^{-1}$ and $\simeq 530$\,km\,s$^{-1}$ when
$\lambda=18^\circ$, and between $\simeq 550$\,km\,s$^{-1}$ and $\simeq
850$\,km\,s$^{-1}$ when $\lambda=162^\circ$. When the solutions are
restricted to those where no Roche-lobe overflow occurs at the time of
the helium star's SN explosion, the minimum kick velocity associated
with $\lambda=18^\circ$ increases slightly to approximately
280\,km\,s$^{-1}$, while the range of admissible kick velocities
associated with $\lambda=162^\circ$ becomes very tightly constrained
to $V_k \approx 640$--680\,km\,s$^{-1}$. The direction of the kick is
again most tightly constrained when $\lambda=162^\circ$: for
$\lambda=18^\circ$ we have $80^\circ \la \theta \la 175^\circ$ and
$|90^\circ-\xi| \ga 5^\circ$, while for $\lambda=162^\circ$ we find
$160^\circ \la \theta \la 175^\circ$ and $5^\circ \la |90^\circ-\xi|
\la 10^\circ$. When the parameter space is restricted to detached
pre-SN binary configurations, $\theta$ and $\xi$ are constrained to
$140^\circ \la \theta \la 175^\circ$ and $5^\circ \la |90^\circ-\xi| \la
15^\circ$ for $\lambda=18^\circ$, and to $\theta \approx
170^\circ$--$175^\circ$ and $|90^\circ-\xi| \approx 5^\circ$ for
$\lambda=162^\circ$.

Hence, for both case~A and~B and for both possible tilt angles
$\lambda$, the formation of PSR\,B1913+16 can only be explained if the
kicks imparted to the second-born NS are ``elevated'' out of the
orbital plane by at least $5^\circ$. In addition, the constraints on
$\theta$ and $\xi$ imply that a large tilt angle of $162^\circ$ cannot
be obtained if the kicks are perpendicular to the pre-SN orbital plane
(i.e. aligned or anti-aligned with the pre-SN orbital angular momentum
vector). Instead, for an unambiguous determination of a large tilt
angle $\lambda$, our results would be more compatible with the
findings of Deshpande, Ramachandran, \& Radhakrishnan (1999) and
Birkel \& Toldr\`a (1997) who argue against a spin-kick correlation.

\begin{figure*}
\resizebox{\hsize}{!}{\includegraphics{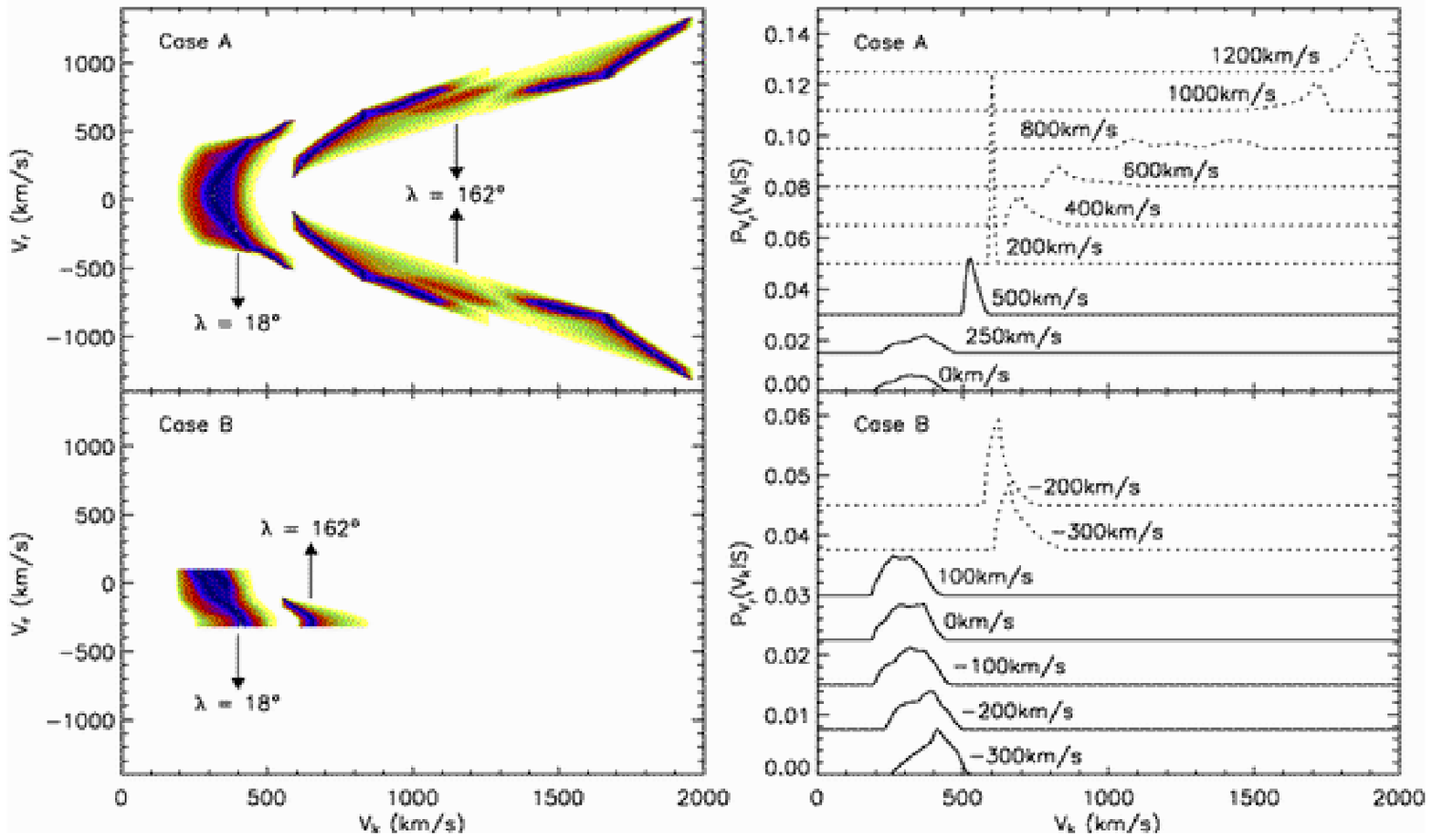}}
\caption{Probability distribution functions of the magnitude of the
kick velocity imparted to PSR\,B1913+16's companion at the time of its
formation, for both $\lambda=18^\circ$ and $\lambda=162^\circ$
(cf. Fig.~\ref{vk1} for more details). In the right-hand panels, the
solid lines correspond to $\lambda=18^\circ$ and the dashed lines to
$\lambda=162^\circ$.}
\label{vk2}
\end{figure*}

The inclusion of pre-SN binary configurations for which stable mass
transfer occurs from a helium star to the first-born NS furthermore
considerably increases the available parameter space for the formation
of PSR\,B1913+16. This is especially true for the prograde case,
$\lambda=18^\circ$, for which the kick direction is no longer required
to be close to the pre-SN orbital plane as was found by Wex et
al. (2000). The conclusion by Wex et al. (2000) that the retrograde
case, $\lambda=162^\circ$, requires more fine-tuning in the kick
direction than the prograde case remains valid even when the
possibility of pre-SN Roche-lobe overflow is taken into account.

The probability distribution functions for the magnitude of the kick
velocity imparted to the second-born NS in PSR\,B1913+16 are shown in
Fig~\ref{vk2} for both $\lambda=18^\circ$ and $\lambda=162^\circ$. The
distributions generally show a single peak at kick velocities which
increase with increasing absolute values of $V_r$. For
$\lambda=18^\circ$, the most probable kick velocity ranges from 
$\simeq 300$\,km\,s$^{-1}$ to $\simeq 600$\,km\,s$^{-1}$, while for
$\lambda=162^\circ$ it ranges from $\simeq 600$\,km\,s$^{-1}$ to
almost $\simeq 2000$\,km\,s$^{-1}$. 

As a test, we also constructed tilt-probability distributions for
PSR\,B1913+16 by omitting the tilt-angle constraints from the
derivation of the admissible parameter space. For case~A, the
distribution functions peak at $\lambda \la 30^\circ$ when $|V_r| \la
400$\,km\,s$^{-1}$ and at $\lambda \ga 150^\circ$ when $|V_r| \ga
900$\,km\,s$^{-1}$. Intermediate absolute values of the radial
velocity yield predictions for the tilt angle which are incompatible
with the observationally derived values. For case~B, the distributions
peak at tilt angles $\lambda \la 30^\circ$ and become vanishingly
small for $\lambda \ga 30^\circ$ over the entire range of possible
$V_r$-values (see Fig.~\ref{pro9}). A large tilt angle of $\lambda =
162^\circ$ can therefore only be explained if the system has crossed
the Galactic plane only once, which would imply that its age is of the
order of {\em only a few Myr}. It is also interesting to note that a
tilt angle of $18^\circ$ would be most favored for radial velocities
around -300\,km\,s$^{-1}$ and +400\,km\,s$^{-1}$, and a tilt angle of
$162^\circ$ for any radial velocity smaller than -900\,km\,s$^{-1}$ or
larger than +900\,km\,s$^{-1}$.

\section{CONCLUSIONS AND DISCUSSION}

In this paper, we have extended the investigation by Willems \&
Kalogera (2004) and have derived a more elaborate and tighter set of
constraints for the progenitor of PSR\,J0737-3039 just before the SN
explosion that forms the second NS as well as for the magnitude and
the direction of the kick imparted to this NS at birth. The additional
constraints stem from the kinematic history of the system which is
derived by tracing the system's motion in the Galaxy backwards in time
and by assuming that its peculiar velocity results entirely from the
second SN explosion. In order to trace the system's motion backwards
in time, we use the scintillation velocity measurements carried out by
Ransom et al. (2004), which yield two velocity components in the plane
perpendicular to the line-of-sight. Since the orientation of the
velocity components is unknown, their orientation $\Omega$ in the
plane perpendicular to the line-of-sight introduces the first free
parameter in the problem. In order to fully describe the system's
space motion, the scintillation velocity components must furthermore
be supplemented with the system's unknown radial velocity $V_r$, the
second free parameter in the problem.

We find that, depending on the values of $\Omega$ and $V_r$, the
system may have crossed the Galactic plane up to two times. We
identify these crossings as possible birth sites of PSR\,J0737-3039
and use the times of the crossings as estimates for the system's age.
It follows that there is a wide range of $\Omega$- and $V_r$-values
for which PSR\,J0737-3039 may be remarkably young ($\la 20$\,Myr). If
the system has crossed the Galactic plane twice, it is at least
20\,Myr old regardless of the values of $\Omega$ and $V_r$.  Our
kinematical analysis furthermore shows that the post-SN peculiar
velocity must have been at least on the order of $\simeq
90$\,km\,s$^{-1}$ and may even have been as high as $\simeq
1200$\,km\,s$^{-1}$ (see Fig.~\ref{vcm1}). This is in contrast to the
work done by Piran \& Shaviv (2004) who used the present proximity of
the system to the Galactic plane to derive a statistical 95\%
confidence upper limit of $\simeq 150$\,km\,s$^{-1}$ on the post-SN
peculiar velocity.

\begin{deluxetable*}{lcccc}
\tabletypesize{\scriptsize}
\tablewidth{0pt}
\tablecolumns{5}
\tablecaption{Summary of the constraints derived for PSR\,J0737-3039,
  PSR\,B1534+12, and PSR\,B1913+16. For PSR\,B1534+12, and
  PSR\,B1913+16 numbers between parentheses correspond to the
  constraints associated with detached pre-SN progenitors. \label{sum}}
\tablehead{ \colhead{Parameter} & 
     \colhead{PSR\,J0737-3039} &
     \colhead{PSR\,B1534+12} &
     \multicolumn{2}{c}{PSR\,B1913+16} \\ 
     \cline{4-5} \\
     \colhead{} &
     \colhead{} &
     \colhead{} &
     \colhead{$\lambda$\tablenotemark{f}$=18^\circ \pm 6^\circ$} &
     \colhead{$\lambda$\tablenotemark{f}$=162^\circ \pm 6^\circ$} 
     } 
\startdata
$M_0$ ($M_\odot$)\tablenotemark{a}    & 2.1--4.7 & 2.1--8.0 (4.7--8.0)  & 2.1--8.0 (4.5--8.0) & 2.1--8.0 (4.5--8.0) \\
$A_0$ ($R_\odot$)\tablenotemark{b}    & 1.2--1.7 & 2.4--4.3 (2.5--4.3)  & 1.1--5.3 (2.5--5.3) & 1.1--5.3 (2.5--5.3) \\
$V_k$ (km\,s$^{-1}$)\tablenotemark{c} & 60--1660 & 15--1350 (230--1350) & 190--600 (300--600) & 580--2000 (700--1250) \\
$\theta$ (degrees)\tablenotemark{d}   & 115--180 &  95-180  (140--180)  & 95--175  (145--175) & 160--175 (165--175) \\
$\xi$ (degrees)\tablenotemark{e}      &  25--155 &  5--175  (55--125)   & 5--85 or 95--175 (75--85 or 95--105) & 75--85 or 95--105 ($\approx$85 or $\approx$95) 
\enddata
\tablenotetext{a}{Pre-SN mass of the last-born NS's helium star progenitor.}
\tablenotetext{b}{Pre-SN orbital separation.}
\tablenotetext{c}{Kick-velocity magnitude.}
\tablenotetext{d}{Angle between the kick direction and the pre-SN
  orbital velocity of the last-born NS's helium star progenitor.} 
\tablenotetext{e}{Angle between the kick direction and the pre-SN
  orbital angular momentum axis.}
\tablenotetext{f}{Misalignment angle between PSR\,B1913+16's spin axis
and the post-SN orbital angular momentum axis.}
\end{deluxetable*}

\begin{figure*}
\begin{center}
\resizebox{5.4cm}{!}{\includegraphics{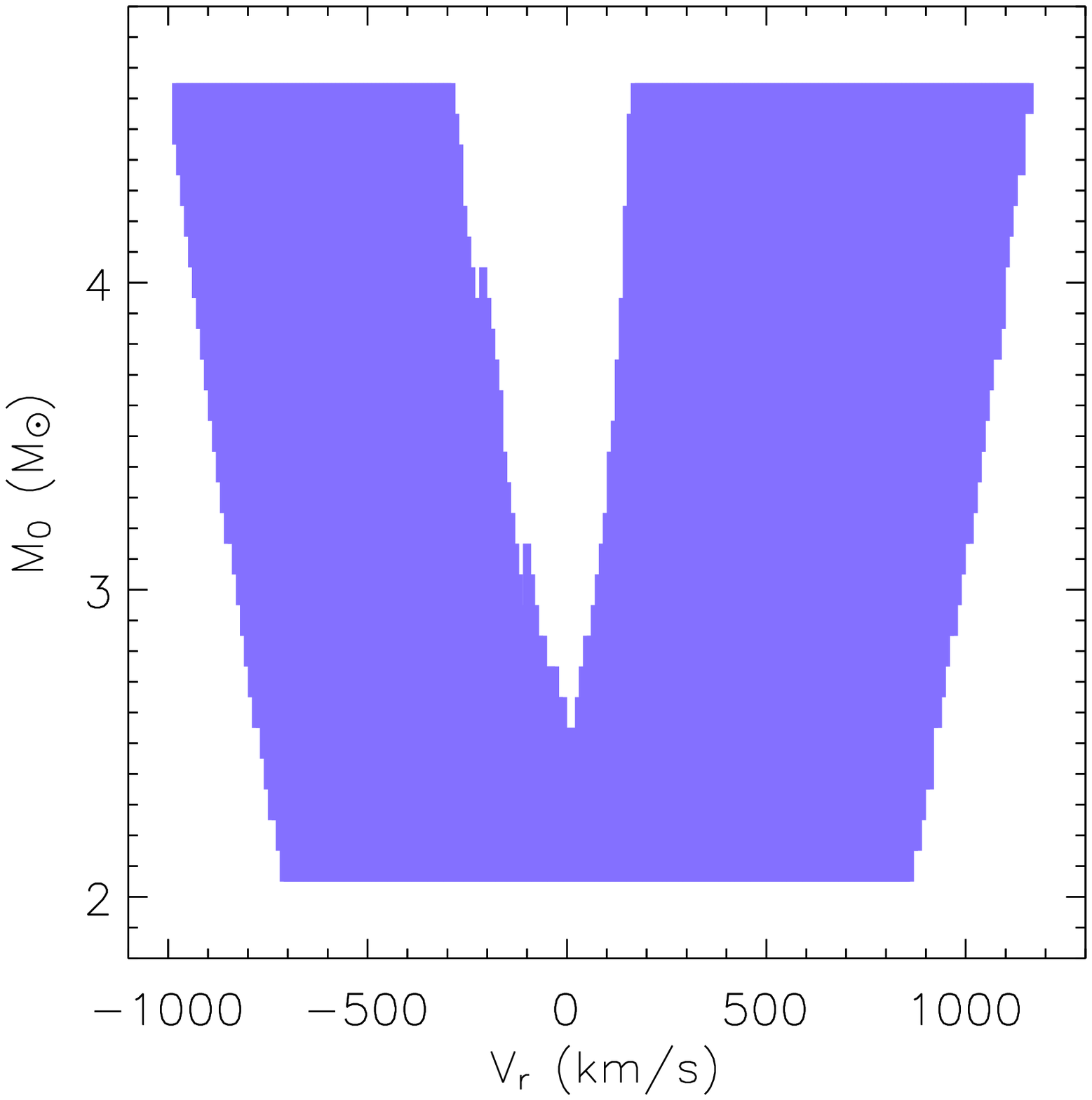}}
\resizebox{5.4cm}{!}{\includegraphics{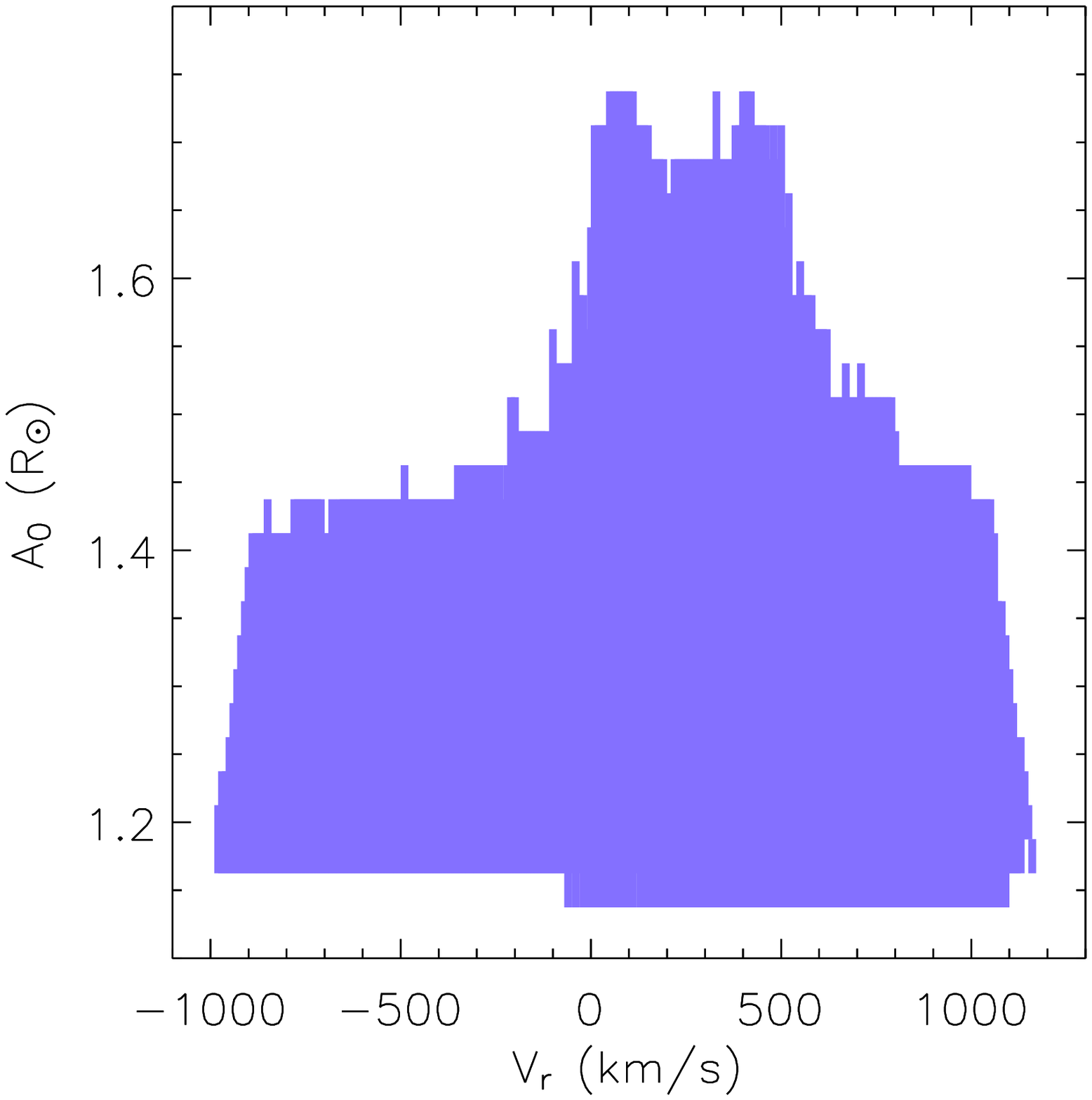}} 
\resizebox{5.4cm}{!}{\includegraphics{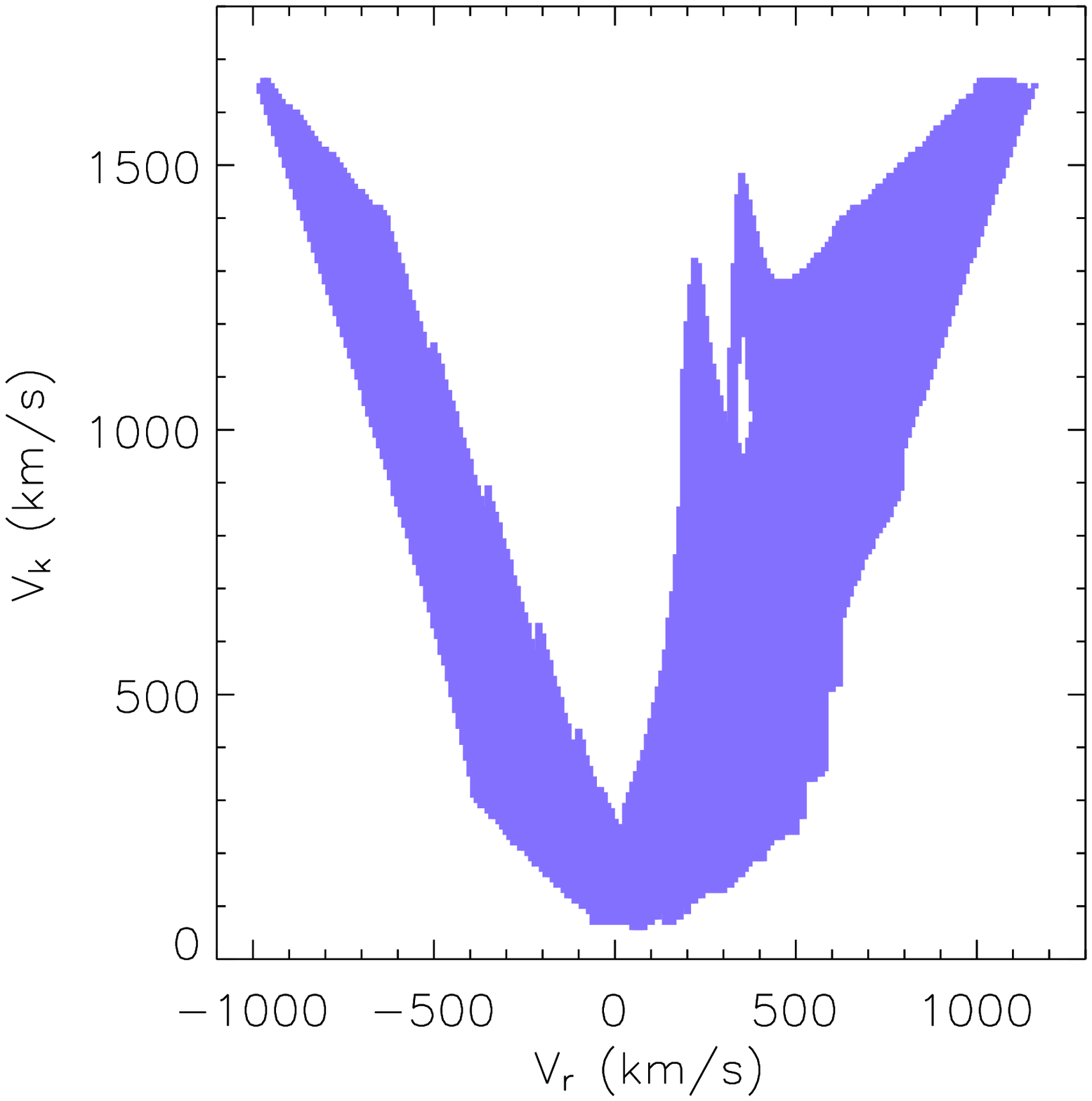}}
\caption{Overview of the constraints on the mass $M_0$ of
PSR\,J0737-3039B's helium star progenitor, the pre-SN orbital
separation $A_0$, and the magnitude $V_k$ of the kick velocity
imparted to pulsar~B at birth, as a function of $V_r$ for all possible
values of $\Omega$.}
\label{overview}
\end{center}
\end{figure*}

Next, we used the age estimates and peculiar velocities associated
with the Galactic plane crossings to impose additional constraints on
the progenitor and formation of PSR\,J0737-3039 besides those already
discussed in Willems \& Kalogera (2004). We find that despite the two
degrees of freedom, the pre-SN and kick parameters as well as the age
become much more constrained for many combinations of $\Omega$ and
$V_r$ (see Figs.~\ref{pro1}--\ref{pro3}). An overview of the
constraints on the mass $M_0$ of pulsar~B's helium star progenitor,
the pre-SN orbital separation $A_0$, and the magnitude $V_k$ of the
kick velocity imparted to pulsar~B at birth as a function of $V_r$ for
all possible values of $\Omega$ is given in Fig.~\ref{overview}. The
absolute limits covering the whole range of admissible $\Omega$-
and $V_r$-values are summarized in Table~\ref{sum}.

We also examined constraints on the kick direction in terms of the
angle $\theta$ between the kick direction and the pre-SN orbital
velocity of pulsar~B's helium star progenitor, as well as the angle
$\xi$ between the kick direction and the pre-SN orbital angular
momentum vector (which could be associated with pulsar B's spin
axis). We find that $\theta$ is always larger than $\simeq 115^\circ$,
while $\xi$ ranges from $\simeq 25^\circ$ to $\simeq 155^\circ$. The
kick imparted to pulsar~B at birth must therefore have been directed
opposite to the orbital motion and cannot have been too closely
aligned with the pre-SN orbital angular momentum axis.

For each combination of $\Omega$ and $V_r$ yielding viable progenitors
for PSR\,J0737-3039, we derived probability distributions for the
magnitude of the kick velocity under the assumption that all kick
directions are equally probable (Fig.~\ref{vk1}). The distribution
functions exhibit a single peak at kick velocities which increase from
$\simeq 100$\,km\,s$^{-1}$ to $\simeq 1600$--1700\,km\,s$^{-1}$ with
increasing absolute values of the radial velocity. This is in contrast
to Willems \& Kalogera (2004) who found a single most probable kick
velocity of $\simeq 150$\,km\,s$^{-1}$. The dependence of the position
of the peak on the radial velocity found here stems from the relation
between the kick magnitude and the post-SN peculiar velocity which
tends to select those kick velocities which are able to explain the
system's motion in the Galaxy (high space velocities cannot be
explained by low kick velocities and vice versa). However, for a given
radial velocity, the range of kick velocities with non-vanishingly
small probabilities found here is always much more constrained than in
Willems \& Kalogera (2004).

Similarly, we derived distribution functions for the misalignment
angle $\lambda$ between pulsar~A's spin axis and the post-SN orbital
angular momentum axis (Fig.~\ref{tilt1}). Tilt angles lower than
$30^\circ$--$50^\circ$ are clearly favored in the cases where (i) the
system has crossed the Galactic plane twice in the past and (ii) the
system crossed the Galactic plane once and has a current radial
velocity of less than $\simeq 500$\,km\,s$^{-1}$ in absolute
value. Higher tilt angles must be associated with one disk crossing
and much higher present-day radial velocities of $\simeq
500$--1200\,km\,s$^{-1}$ in absolute value. Tilt angles close to
$90^\circ$ are furthermore strongly disfavored for any radial velocity
$V_r$.  Our probability distributions are thus compatible with the
tilt-angle predictions of $16^\circ \pm 10^\circ$ and $164^\circ \pm
10^\circ$ made by Jenet \& Ransom (2004) and with the spin- magnetic
dipole misalignment angle derived by Demorest et al. (2004) which
rules out Jenet \& Ransom's alternative solutions of $82^\circ \pm
16^\circ$ or $98^\circ \pm 16^\circ$.

In view of recent claims of alignment of NS kicks with NS spin axes
(e.g. Romani 2004), we also considered kick-velocity and tilt-angle
distributions for polar kicks where the kick direction is confined
within two oppositely directed cones with an opening angle of
$30^\circ$ and with axes parallel to the pre-SN orbital angular
momentum axis (i.e. $\xi \le 30^\circ$).  The confinement of the kick
directions strongly reduces the range of radial velocities for which
viable progenitors for PSR\,J0737-3039 may be found. Correspondingly,
the range of most probable kick velocities shrinks to $\simeq
200$--550\,km\,s$^{-1}$, and the range of most probable tilt angles to
$15^\circ$--$45^\circ$ (see Figs.~\ref{vk1b} and~\ref{tilt1b}).

Progenitor constraints and isotropic kick-velocity and spin-tilt
distributions were also derived for PSR\,B1534+12 and PSR\,B1913+16.
Both systems have a measured proper motion with a known direction in
the plane of the sky, so that the derived constraints and probability
distributions only depend on the unknown radial velocity $V_r$. For
PSR\,B1913+16, the knowledge of the tilt angle $\lambda$ (Kramer 1998;
Weisberg \& Taylor 2002) furthermore puts additional constraints on
the possible progenitor systems, which are not yet available for
PSR\,J0737-3039 and PSR\,B1534+12. 

The absolute limits on the pre-SN binary and kick parameters of the
two systems are summarized in Table~\ref{sum} (the dependency of the
constraints on $V_r$ is shown Figs.~\ref{pro10}
and~\ref{pro9}). Contrary to PSR\,J0737-3039, the progenitors of these
two systems were not necessarily undergoing a mass-transfer phase at
the time the second NS was born. The helium star masses $M_0$, pre-SN
orbital separations $A_0$, and kick velocities $V_k$ associated with
detached progenitors are, however, always much more constrained than
those associated with the full set of detached and semi-detached
progenitors. In particular, the possibility of pre-SN Roche-lobe
overflow considerably relaxes the constraints on the progenitor of
PSR\,B1913+16 derived by Wex et al. (2000).

The constraints on the direction of the kick imparted to the
second-born NS in PSR\,B1534+12 and PSR\,B1913+16 yield very similar
conclusions as for PSR\,J0737-3039: the kicks are generally directed
opposite to the orbital motion and cannot have been too closely
aligned with the pre-SN orbital angular momentum axis.

The kick-velocity and spin-tilt distributions for PSR\,B1534+12 and
PSR\,B1913+16 (Figs.~\ref{vk4}, \ref{tilt3} and~\ref{vk2}) also show
qualitatively the same behavior as for PSR\,J0737-3039. The most
probable kick velocity depends on the unknown radial velocity and,
depending on the value of $V_r$, can be anywhere between the derived
lower and upper limits for the kick imparted to the second-born
NS. Tilt angles smaller than $30^\circ$--$50^\circ$ are again favored
when (i) the system crossed the Galactic plane more than once and (ii)
the system crossed the Galactic plane once and has a current radial
velocity smaller than $\simeq 250$\,km\,s$^{-1}$ in absolute value for
PSR\,B1534+12 and smaller than $\simeq 500$\,km\,s$^{-1}$ in absolute
value for PSR\,B1913+16.

For conclusion, we point out that the set of observational
measurements and constraints available for double neutron stars
systems are sufficient to greatly limit the properties of the
progenitors. As a follow-up investigation it will be therefore
interesting to compare these strongly constrained progenitor
properties to synthesis studies of binary populations and double
neutron star formation.

\acknowledgments 
We are grateful to Scott Ransom for sharing the scintillation velocity
measurements for PSR\,J0737-3039 prior to publication; to Ingrid
Stairs for pointing out the correct measurements of the proper motion
of PSR\,B1534+12 before the publication of the paper; to Steve
Thorsett for helping us identify a sign error in the Galactic motion
calculations for PSR\,B1534+12; and to Laura Blecha for sharing the
code used to trace the motion of the DNS binaries in the Galactic
potential. We also thank the referee, Philipp Podsiadlowski, for his
useful suggestions and remarks which helped to improve the paper. The
organizers of the Aspen Winter Conference on Binary Radio Pulsars are
acknowledged for putting together a stimulating conference during
which this research was initiated. This work is partially supported by
a NSF Gravitational Physics grant, a David and Lucile Packard
Foundation Fellowship in Science and Engineering grant, and NASA ATP
grant NAG5-13236 to VK.

\end{document}